\documentclass[pra,amsmath,showpacs,twocolumn,showkeys,letterpaper]{revtex4}

\usepackage{times}
\usepackage{graphicx}
\usepackage{marvosym}
\usepackage{dsfont}
\usepackage{bm}

\usepackage{textcomp} 
\usepackage{wasysym}

\newcommand{\eq}[1]{Eq.~(\ref{#1})}
\newcommand{\bra}[1]{\left\langle#1\right|}
\newcommand{\ket}[1]{\left|#1\right\rangle}

\newcommand{\scp}[2]{\langle #1\!\mid\!#2\rangle}
\newcommand{\ave}[1]{\langle#1\rangle}

\newcommand{\ie}{i.\thinspace{}e., }

\newcommand{\bxi}{{\boldsymbol{\xi}}}
\newcommand{\bp}{{\mathbf{p}}}

\newcommand{\aop}[1]{{\hat{a}^{\vphantom\dag}(#1)}}
\newcommand{\aopd}[1]{{\hat{a}^{\dag}(#1)}}

\newcommand{\brho}{\boldsymbol{\rho}}
\newcommand{\bpi}{\boldsymbol{\pi}}

\newcommand{\brhob}{\Bar{\boldsymbol{\rho}}}

\newcommand{\bpib}{\bar{\boldsymbol{\pi}}}

\newcommand{\rhob}{\bar{\rho}}

\newcommand{\bpbp}{\bar{\mathbf{p}}'}

\newcommand{\br}{\mathbf{r}}
\newcommand{\brb}{\bar{\mathbf{r}}}
\newcommand{\brbp}{\bar{\mathbf{r}}'}
\newcommand{\dbrbp}{\dot{\bar{\mathbf{r}}}'}

\newcommand{\bro}{\hat{\mathbf{r}}}
\newcommand{\bpo}{\hat{\mathbf{p}}}

\newcommand{\bl}{\mathbf{l}}

\newcommand{\Rop}{{\Hat{\mathbf{R}}}}
\newcommand{\Pop}{{\Hat{\mathbf{P}}}}
\newcommand{\Lop}{{\Hat{\mathbf{L}}}}
\newcommand{\Nop}{{\Hat{N}}}

\newcommand{\Rc}{{\boldsymbol{\mathcal{R}}}}
\newcommand{\Reas}{\boldsymbol{\Re}}
\newcommand{\Reasb}{\bar{\boldsymbol{\Re}}}

\newcommand{\Pc}{{\boldsymbol{\mathcal{P}}}}
\newcommand{\Qc}{{\mathcal{Q}}}
\newcommand{\Qcb}{{\boldsymbol{\mathcal{Q}}}}
\newcommand{\Sc}{{\mathcal{S}}}
\newcommand{\Oeas}{{\boldsymbol{\Omega}_{{\tiny \earth}}}}
\newcommand{\Oeasb}{{\bar{\boldsymbol{\Omega}}_{{\tiny \earth}}}}
\newcommand{\Oe}{{\Omega_{{\tiny \earth}}}}
\newcommand{\Oet}{{\Omega_{{\tiny \earth}}t}}

\newcommand{\Uopdag}{{\hat{U}^{\dagger}}}
\newcommand{\Uop}{{\hat{U}^{\vphantom\dagger}}}
\newcommand{\URdag}{{\hat{U}_\mathcal{R}^{\dagger}}}
\newcommand{\UR}{{\hat{U}_\mathcal{R}^{\vphantom \dagger}}}
\newcommand{\UPdag}{{\hat{U}_\mathcal{P}^{\dagger}}}
\newcommand{\UP}{{\hat{U}_\mathcal{P}^{\vphantom \dagger}}}
\newcommand{\UOdag}{{\hat{U}_\Qc^{\dagger}}}
\newcommand{\UO}{{\hat{U}_{\Qc}^{\vphantom \dagger}}}

\newcommand{\UNdag}{{\hat{U}_\Sc^{\dagger}}}
\newcommand{\UN}{{\hat{U}_\Sc^{\vphantom \dagger}}}

\newcommand{\brp}{\mathbf{r}^\prime}
\newcommand{\bna}[2]{\frac{\partial #2}{\partial #1 }}

\newcommand{\eas}{{\small \earth}}

\begin{document}
\title{Dropping cold quantum gases on Earth over long times and large
  distances} \author{G.~Nandi}
 \email{gerrit.nandi@uni-ulm.de}
\author{R.~Walser}
\author{E.~Kajari}
\author{W.~P.~Schleich}
\affiliation{Institut f\"ur Quantenphysik, Universit\"at Ulm,
  D-89069 Ulm, Germany}
\date{\today} 

\begin{abstract}
  We describe the non-relativistic time evolution of an ultra-cold degenerate
  quantum gas (bosons/fermions) falling in Earth's gravity during long times
  (10 sec) and over large distances (100 m). This models a drop tower
  experiment that is currently performed by the QUANTUS collaboration at ZARM
  (Bremen, Germany).  Starting from the classical mechanics of the drop
  capsule and a single particle trapped within, we develop the quantum field
  theoretical description for this experimental situation in an inertial
  frame, the corotating frame of the Earth, as well as the comoving frame of
  the drop capsule.  Suitable transformations eliminate non-inertial forces,
  provided all external potentials (trap, gravity) can be approximated with a
  second order Taylor expansion around the instantaneous trap center. This is
  an excellent assumption and the harmonic potential theorem applies.  As an
  application, we study the quantum dynamics of a cigar-shaped Bose-Einstein
  condensate in the Gross-Pitaevskii mean-field approximation.  Due to the
  instantaneous transformation to the rest-frame of the superfluid wave
  packet, the long-distance drop (100m) can be studied easily on a numerical
  grid.
\end{abstract}

\maketitle

\section{Introduction}

Dropping toys on the floor is one of the earliest childhood experiences and
remains a source of endless joy.  Modern physics, as we know it today, is also
based on the consequent pursuit of this naive amazement about the gravitational
attraction between material bodies.  The contributions of Galilei, Newton and 
Einstein to the understanding of the free fall have fundamentally changed
the way we understand modern physics.

Nowadays, the gravitational field of the Earth is under more intense scrutiny
than ever before.  One research branch focuses on the complex geodynamics of
the classical gravitational field of the Earth. From tide movements of the
oceans and the atmospheric mass flows to minute wobbling of the instantaneous
Earth rotation axes due to liquid core motion, all such effects become
measurable.  This is either done with ground based gravitometers that measure
the time-dependent local acceleration $g$ \cite{peters99}, Earth's rotation
\cite{werner79,packard971,lasergyro,schleich84,wettzell,durfee05}, or in space where
satellite-based geodesic measurements \cite{CHAMP,GOCE,GRACE} put tighter
limits on the higher multipole moments of the gravitational potential
\cite{multipoleearth}.  Another research direction focuses more on the
fundamental aspects of gravity which follows from general relativity, for
example the current measurement of the gravito-magnetic field with orbiting
gyroscopes by the "Gravity probe B" experiment \cite{gravityb}.
This gravitational science on Earth and in space connects different
branches of physics \textendash\, on the atomic, macroscopic and cosmological
level. It is vigorously pursued by American, Russian, European, as well as
Chinese space agencies.

The discovery of Bose-Einstein condensation \cite{pethick02,stringaribuch} and
fermionic superfluidity \cite{regal04,jochim03,zwierlein04} in dilute atomic
vapors has introduced new members to the family of superfluid condensed
matter systems.  Outstanding features of atomic vapors are that they exists
at the lowest possible temperatures, \ie on the nano- and pico-Kelvin scale,
and their dynamic properties can be engineered externally.

These unique features are now combined in an experiment to study Bose-Einstein
condensates in $\mu$-gravity. At the drop tower facility of ZARM (Center of
Applied Space Technology and Microgravity, University of Bremen, Germany), the
QUANTUS collaboration \cite{quantus} focuses currently on the implementation of
a fully selfcontained $^{87}$Rb-BEC experiment that fits into a small drop
capsule and falls repeatedly over a distance of 100 m. This is performed
inside an evacuated drop tower tube and results in a residual gravitational
acceleration of $\Delta g/g=10^{-6}$. A status report of the experiment is
given in Ref.~\cite{becmugrav06}.

The physics that can be explored with the falling BEC naturally splits into two
topics.  First, one can consider the extension of atom interferometry
\cite{berman,raselertmer05,canuel06,dubetsky06} for high-precision inertial 
and rotation sensing.
Due to the long unperturbed free-fall times of up to 10 seconds, it can be
expected that precision of measurements can be improved accordingly.  Second,
fundamental questions regarding the quantum nature of degenerate gases can be
studied.  Due to the possibility of decreasing the trapping potential
significantly in a micro-gravity environment without the need of external
levitation fields to compensate for gravity, lower ground state energies than
ever should be accessible and the pico-Kelvin physics could be entered
\cite{leanhardtscience03}.  In the resulting ultra-large condensates
(10mm), it is possible to gain absolute control of the macroscopic matter-wave
as optical readout and manipulation can be performed with very high relative
spatial resolution.  Furthermore, a long-time unconstraint expansion of a BEC
allows for a measurement of the macroscopic wave function, its higher
correlations \cite{blochnature05,westbrook06} and probes the very concept of long-range order.
Exact symmetries of the system such as the Kohn mode
\cite{kohn,dobson,birula02} or the breathing mode \cite{pitaevskii97} can be
studied in unprecedented detail.

The theoretical description of any of the previously mentioned topics requires
a detailed modeling of the launch procedure, the subsequent motion of the drop
capsule, the dynamics of the comoving condensate and its time-dependent
trapping geometry.  This requires the numerical solution of the
Gross-Pitaevskii equation for the semi-classical field amplitude.  If we
describe it on a numerical grid, which rests in the comoving frame of the BEC
wave packet, then this task becomes most simple. Obviously, there is the need to
describe the mapping between this particular non-inertial frame and the other
possible reference frames used for observation.  These considerations can be
extended to the quantum depletion and the pair correlation function
\cite{walser599,kramer06}.

This article is organized as follows: In Section II we shortly review the
classical physics of a single harmonically trapped particle falling within the
drop capsule.  Section III is dedicated to the quantum mechanical description
of a many-particle system, whether bosons or fermions, trapped in a harmonic
potential and falling within a capsule in the gravitational field. We consider
the unitary representations of the coordinate transformations on many-particle
Fock space. We derive a harmonic potential theorem \cite{dobson} and obtain
the many-particle quantum evolution in the reference frame of the capsule,
which decouples from the free-fall motion according to the equivalence
principle.  As an application, we study the quantum dynamics of a
Bose-Einstein condensate (BEC) in the Gross-Pitaevskii (GP) mean-field
approximation. Due to the instantaneous transformation to the rest-frame of
the capsule, the long-distance drop (100m) can be studied easily on a
numerical grid.  In the previous sections, we have deliberately disregarded
the rotation of the Earth. This is rectified in Section IV.  In there, we
obtain the main result of the article, i.e. the complete quantum mechanical
description of the drop tower experiment, as well as the transformation rules
for observables between the tower- and the drop capsule frame.  In Section V,
we extend the theory for structureless particles to two-level atoms, which can
be coupled via an external, off-resonant laser. In the special case of equal
scattering lengths (e.g. $^{87}$Rb), we show that the internal dynamics
decouples from the external motion.

\section{Classical physics of a particle falling within the drop capsule}

In this section, we review the classical physics of a single harmonically
trapped particle in a free-fall experiment.  Therefore, we need to understand
the classical motion of the drop capsule first.

\subsection{The drop capsule}
\subsubsection{Experimental configuration}
The drop capsule is a cylindrical container that houses the complete setup
studied in the release experiment \cite{zarm}. It is shaped like a projectile,
with a diameter of $80$ cm, a height of $2.4$ m and a gross mass $\mathcal{M}$
of up to $500$ kg.  Either, it is lifted $110$ m to the top of the
depressurized tower tube ($<10$ Pa) and released to fall freely ($4.74$ s), or
it can be shot up from the ground with a powerful air-gun driven catapult,
thus doubling the time available for ballistic motion. On impact, it is
decelerated smoothly with a container full of polystyrene pellets. The same
experiment can be repeated up to three times per day.

\begin{figure}[h]
\begin{center}
\includegraphics[width=\columnwidth]{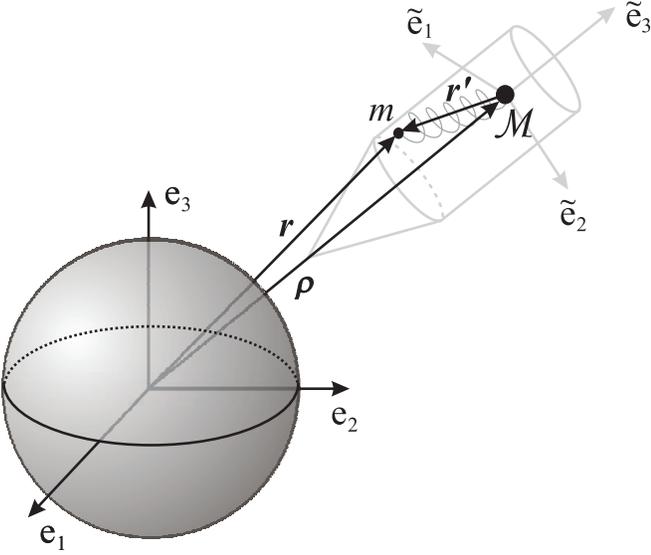}
\caption{
  A harmonically trapped particle with position $\br$ and mass $m$ falls
  towards the Earth in a drop capsule with position $\brho$ and mass
  $\mathcal{M}$. The center-of-mass of the drop capsule $\brho$ and its
  major axes coincide also with the origin of the harmonic trap and its
  principal axes $\mathbf{\widetilde{e}}_i$.}
\label{coord1}
\end{center}
\end{figure}

Irrespective of which specific launch procedure is chosen, it is clearly
necessary to briefly review the basic Newtonian physics of the drop capsule.
In the simplest scenario, we may safely disregard any tumbling micro-motion of
the capsule along its flight path, causing gyro-mechanical effects. Thus, we
will model the capsule solely by its center-of-mass coordinate
$\brho=\sum_{i=1}^3\rho^i \, \mathbf{e}_i.$ We assume that a Cartesian
coordinate system, fixed to the center of the Earth, is a good inertial
reference frame with basis vectors denoted by
$\{\mathbf{e}_1,\mathbf{e}_2,\mathbf{e}_3\}$.  This configuration is shown in
Fig.~\ref{coord1}.  For the moment, we will also ``freeze'' the rotation of
the Earth and consider those effects explicitly in
Sec.~\ref{rotframe}.

In general, the gravitational potential of the Earth $V_g$ is not
circular-symmetric, nor stationary. From a geophysical viewpoint, Earth
resembles a drop of a viscous fluid. During the past eons, it evolved into an
oblate ellipsoid due to its rotation.  Consequently, the local gravitational
acceleration $\bm{g}$ is orthogonal to the surface of the Earth, but does not
point towards the center, except for the equator and the poles \cite{greiner}.
Even today, the geodynamic activity has not subsided but remains noticeable in
the form of tidal oscillations and liquid core wobble.  Thus, for the purpose
of modeling high-precision drop experiments, one needs to consider the general
expression for the gravitational potential
\begin{equation}
\label{grav}
V_g(\brho,t)=-G\int_E\text{d}^3x\frac{m_{\eas}(\mathbf{x},t)}{
|\brho-\mathbf{x}|},
\end{equation}
where $G=6.6742\times 10^{-11} \ \mbox{m}^3\, \mbox{s}^{-2} \,
\mbox{kg}^{-1}$ represents the gravitational constant \cite{codata} and
$m_{\eas}$ denotes the mass density of the Earth.  Currently, experimental
data in the form of a multipole expansion to the 360$^{th}$ degree is
available \cite{multipoleearth} and more satellite based geodetic measurements
are on the way \cite{GOCE,GRACE,CHAMP}.  Despite this remarkable precision in
the gravitometric data, it remains nevertheless true that the monopole is the
dominant contribution to the gravitational potential. It is proportional to
the standard gravitational constant, i. e. the product of the gravitational
constant and the total mass of the Earth
$G\,\int_E\text{d}^3x\,m_{\eas}(\mathbf{x},t)= 3.986\times 10^{5}\ 
\mbox{km}^{3} \, \mbox{s}^{-2}$.
\subsubsection{Hamiltonian dynamics}
In considering the mass $\mathcal{M}$ of the drop capsule, it is clear that
classical mechanics rules its dynamics. The succinct formulation of analytical
mechanics follows from Hamilton's principle, which demands that the trajectory
defined by position $\brho$ and orientation of the capsule
$\mathbf{n}$, extremizes the classical action
\begin{gather}
  \label{action}
  S_c[\brho,\mathbf{n}]=\int_{t_0}^{t} \text{d}t' \,
  L_c[\brho(t'),\dot{\brho}(t'),\dot{\mathbf{n}}(t'),t'],
\end{gather}
subject to appropriate boundary conditions \cite{goldstein}.  The Lagrangian
for the drop capsule $L_c$ with respect to an inertial frame is given by
\begin{gather}
  \label{lag1}
  L_c[\brho,\dot{\brho},\dot{\mathbf{{n}}},t]=
  \mathcal{M}\left(\frac{\dot{\brho}^2}{2}
    -V_g(\brho,t)\right)+\frac{1}{2}\dot{\mathbf{n}}\,\mathcal{I}
  \,\dot{\mathbf{n}}.
\end{gather}
In there, we introduced the tensor of inertia $\mathcal{I}$.  The rate of
change of $\mathbf{n}$ corresponds to the angular velocity or spinning of the
capsule.  As mentioned in the introduction, we will disregard possible
torques, which arise due to weak gravity inhomogeneities.  Thus, the internal
angular momentum of the drop capsule is conserved
$\mathcal{I}\dot{\mathbf{n}}=\text{const.}$, in other words $\mathbf{n}$ is a
cyclic variable in the Lagrangian Eq.~(\ref{lag1}).
To simplify the following discussion, we will also drop the energy offset
$\dot{\mathbf{n}}\,\mathcal{I}\,\dot{\mathbf{n}}/2$, which has no
dynamical consequences.

The global extremum of the action $S_c$ is attained when the trajectory
satisfies locally the Euler-Lagrange equations
\begin{gather}
  \label{EuLa}
  \frac{d}{dt}\bna{\dot{\brho}}{L_c}= \bna{\brho}{L_c}.
\end{gather}
The transition from the Lagrangian to the Hamiltonian
formulation is based on the introduction of the canonical momenta
\begin{align}
  \bpi(\dot{\brho})&=\bna{\dot{\brho}}{L_c}=\mathcal{M} \dot{\brho}.
\end{align}
Provided the canonical momenta can be expressed uniquely via velocities, then
the Legendre transformation of the Lagrangian is invertible,
$\text{Det}(\partial^2 L_c/\partial \dot{\rho}_i \partial \dot{\rho}_j)\neq
0$. It defines a Hamiltonian function in phase space
\begin{align}
  H_c(\brho,\bpi,t)&=\bpi \dot{\brho}(\bpi)
-L_c,
\end{align}
and the Hamiltonian equations of motion represent the dynamics
\begin{alignat}{2}
  \dot{\brho}&=\bna{\bpi}{H_c}, &\qquad
  \dot{\bpi}&=-\bna{\brho}{H_c}.
\intertext{
In particular, applying analytical mechanics to the drop capsule in vacuum,
falling in gravity, we find} 
\label{dyncap}
  \dot{\brho}&=\frac{\bpi}{\mathcal{M}}, &\qquad
  \dot{\bpi}&=-\mathcal{M}\bna{\brho}{} V_g(\brho,t) ,
\end{alignat}
or simply Newton's equation
\begin{gather}
  \label{force}
  \ddot{\brho}=-\bna{\brho}{}V_g(\brho,t).
\end{gather}

\subsection{A single classical particle trapped in the drop capsule}
\label{secsingle}

For our considerations the inertial frame of reference as well as the comoving
frame of the drop capsule are of significance. Therefore, we will briefly
discuss them in the following sections.

\subsubsection{Inertial frame}
The situation considered in Fig.~\ref{coord1} is almost analogous to
Newton's proverbial apple dropping in the falling elevator. However, in
addition to the gravitational acceleration, our particle with coordinate $\br$
experiences a linear trapping force, possibly time-dependent. This force is
derived from a harmonic oscillator potential
\begin{align}
  \label{eq:trappot}
  V_t(\bxi,t)=&\frac{1}{2}\, (\bxi \otimes \bxi)\cdot v_t^{(2)}(t),\\
  v_t^{(2)}(t)=&\sum_{i=1}^3  \omega_{i}^2(t)\
  \mathbf{\widetilde{e}}_{i}\otimes \mathbf{\widetilde{e}}_{i},
\end{align}
which is tied to the center-of-mass of the drop capsule $\brho$ and is rigidly
aligned along the symmetry axes of the capsule $\{\mathbf{\widetilde{e}}_1,
\mathbf{\widetilde{e}}_2,\mathbf{\widetilde{e}}_3\}$.  In general, the
potential can be anisotropic with time-dependent trap frequencies
$\omega_i(t)$. Both information is incorporated in the definition of the
symmetric tensor $v_t^{(2)}(t)$.

As before, the equation of motion for the falling, trapped particle follows
straight from the Lagrangian of the drop capsule, Eq.~(\ref{lag1}), by adding
the corresponding Lagrangian of the trapped particle
\begin{gather}
\label{lagcappluspart}
  L[\brho,\dot{\brho},\br, \dot{\br},t]=
  L_c[\brho,\dot{\brho},t]+ L_{sp}[\br,\dot{\br},t],\\
  \label{lagtrap}
  L_{sp}[\br,\dot{\br},t]= m\left(  
    \frac{\dot{\br}^2}{2} 
    -V_t(\br-\brho(t),t)-V_g(\br,t) \right).
\end{gather}
It almost goes without saying that the back-action of the particle on the drop
capsule is negligible, since the mass of the drop capsule $\mathcal{M}\gg m$
is much larger than the mass $m$ of the atomic particle.

In the inertial frame, the atomic canonical momentum $\bp$ is identical with
the kinetic momentum 
\begin{gather}
\bp=\bna{\dot{\br}}{L_{sp}}=m \dot{\br}.
\end{gather} 
Thus, the Hamiltonian function for the single trapped particle is obtained
immediately as
\begin{equation}
\begin{aligned}
  \label{htrap}
  H&_{sp}(\br,\bp,t)=\bp \dot{\br}-L_{sp}\\
  &=\frac{\bp^2}{2 m}+m [V_{{t}}\left(\br-\brho(t),t\right)+V_g(\br,t)]
\end{aligned}
\end{equation}
and the equations of motion read as
\begin{gather}
\dot{\br}=\frac{\bp}{m}, \\
\dot{\bp}=-m\frac{\partial}{\partial \br}[V_{{t}}\left(\br-\brho(t),t\right)+V_g(\br,t)].
\end{gather}
This yields Newton's equation for the particle coordinates $\br$
\begin{gather} 
  \ddot{\br}=-\bna{\br}{} \left[ V_t\left(\br-\brho(t),t\right)+V_g(\br,t)
  \right].
  \label{classicho}
\end{gather}
\subsubsection{Comoving frame}
The particle can deviate only a tiny distance 
from the center of the drop capsule. It is of the order of $1$~cm, hence much
less than the radius of the Earth.  Consequently, it is
prudent to make a coordinate transformation to the comoving, accelerated
frame and also to introduce relative velocities
\begin{align}
\label{trans}
  \br'&=\br-\brho(t),\qquad
\dot{\br}'=\dot{\br}-\dot{\brho}.
\end{align}
The Lagrangian of the particle now reads
\begin{gather}
  \label{lagtrapII}
  L_{sp}'[\brp,\dot{\br}',t]= m\left(\frac{(\dot{\br}'+\dot{\brho})^2}{2}
    -V_t(\brp,t)-V_g(\br'+\brho,t) \right)
\end{gather}
and induces a canonical momentum as
\begin{gather}
\bp'=\frac{\partial  L_{sp}'}{\partial \dot{\br}'}=m(\dot{\br}'+\dot{\brho}).
\end{gather}
Now, the Hamiltonian in the accelerated frame is given by
\begin{equation}
\begin{aligned}
  \label{htrapaccel}
  H_{sp}'(&\brp,\bp',t)=\bp' \dot{\br}'-L_{sp}'\\
  &=\frac{\bp'^2}{2 m}+
  m [V_{{t}}\left(\brp,t\right)+V_g(\brp+\brho,t)]-\bp'\dot{\brho}.
\end{aligned}
\end{equation}
and leads to the following equations of motion
\begin{alignat}{2}
\label{hameqaccelframe}
  \dot{\br}'&=\bna{\bp'}{H_{sp}'}=\frac{\bp'}{m}-\dot{\brho},\\
  \dot{\bp}'&=-\bna{\brp}{H_{sp}'}=-m\bna{\brp}{} 
[V_{{t}}\left(\brp,t\right)+V_g(\brp+\brho,t)],
\end{alignat}
or Newton's equation
\begin{alignat}{2}
\label{newtonaccelframe}
  \ddot{\br}'&=-\bna{\brp}{} 
\left[V_{{t}}\left(\brp,t\right)+V_g(\brp+\brho,t)\right]-\ddot{\brho}.
\end{alignat}

Due to the smallness of the possible deviation of the trapped particle from
the trap center, we can safely expand the gravitational potential at each
instant in a Taylor series along the trajectory of the drop capsule $\brho(t)$,
\begin{equation}
\begin{gathered}
  \label{taylorgrav}
  V_g(\mathbf{r'}+\brho,t)=[1+\mathbf{r'}\,\bna{\brho}{}+
  \frac{1}{2!}(\mathbf{r'}\,\bna{\brho}{})^2+\ldots]\,V_g(\brho,t)\\
  =V_g(\brho,t)+V_g^{(1)}(\br',\brho,t)+V_g^{(2)}(\br',\brho,t)\\
  +\delta V_g^{(3)}(\br',\brho,t).
\end{gathered}  
\end{equation}
This expansion defines the gradient field and a linear potential contribution,
\begin{gather}
\label{grad}
\begin{aligned}
\mathbf{v}_g^{(1)}(\brho,t)&=\bna{\brho}{}\,V_g(\brho,t),\\
V_g^{(1)}(\br',\brho,t)&=\br'\ \mathbf{v}_g^{(1)}(\brho,t),
\end{aligned}
\end{gather}
the symmetric Hessian tensor and its complete contraction into a quadratic
potential energy
\begin{gather}
  \label{hess}
  \begin{aligned}
    v_g^{(2)}(\brho,t)&=(\bna{\brho}{}\otimes \bna{\brho}{}) \ V_g(\brho,t),\\
    V_g^{(2)}(\br',\brho,t)&=\frac{1}{2}\,(\br'\otimes \br') \cdot
    v_g^{(2)}(\brho,t),
  \end{aligned}
\end{gather}
as well as a residual correction $\delta V_g^{(3)},$ given in
Appendix~\ref{apptay}. Due to its smallness, we will tacitly disregard
corrections of this order in all of the following considerations.
\subsubsection{Pre-release dynamics}
\label{pre}
Prior to the release at $t_0$, the capsule is statically attached to the top
of the tower, i.e. $\dot{\brho}(t< t_0)=0$, at some hight $\brho_0$. Please note that
the vanishing velocity implies that we have deliberately ``frozen'' the
rotation of the Earth. This is rectified in Sec.~\ref{rotframe}.  Then, we can
approximate Newton's equation \eq{newtonaccelframe} as
  \begin{align}
    \label{resclass}
    \ddot{\br}'&= -\bna{\brp}{} \tilde{V}_t(\brp,\brho_0,t)
    -\mathbf{v}^{(1)}_g(\brho_0,t),\nonumber\\
    &=-\, \tilde{v}^{(2)}_t(\brho_0,t)\, (\brp-\brp_{eq}(t)).
  \end{align}
  The equilibrium position $\brp_{eq}$ reflects the gravitational sag of the
  particle with respect to the trap center $\brho_0$ at some instant and is
  given by
\begin{gather}
  \tilde{v}^{(2)}_t(\brho_0,t)\, \brp_{eq}(t)= 
-\mathbf{v}^{(1)}_g(\brho_0,t).
\end{gather}
Presumably, its time dependence is very slow.  Furthermore, we have gathered
all quadratic potential energy contributions into a renormalized harmonic
trapping potential as
\begin{equation} 
  \label{eq:trappotmod}
  \begin{gathered}
    \tilde{v}^{(2)}_t(\brho,t)= v_t^{(2)}(t)+v_g^{(2)}(\brho,t),\\
    \tilde{V}_t(\bxi,\brho,t)=\frac{1}{2}\,(\bxi \otimes \bxi)\cdot  
    \tilde{v}^{(2)}_t(\brho,t).
  \end{gathered}
\end{equation}
It is interesting to note that the trapping force is weakened in the
direction of gravity as the curvature coefficients $v_g^{(2)}$ are negative.
\subsubsection{Post-release dynamics}
After the capsule is released, \ie for times $t> t_0$, it falls essentially
according to \eq{force}. Due to the previously mentioned sag in the direction
of gravity, the equilibrated atomic particle will be in a position
$\br'(t_0)=\brp_{eq}(t_0)$ and initiate harmonic oscillations about the
instantaneous trap center $\brho(t)$
\begin{align}
  \label{resclasspost}
  \ddot{\br}'&= -\bna{\br'}{}\
  \tilde{V}_t(\br',\brho(t),t)=-\tilde{v}^{(2)}_t(\brho(t),t)\, \br'.
\end{align}

\subsubsection{Hamiltonian formulation and canonical transformations}
\label{hamform1}
In order to obtain the transformed single-particle Hamilton function
$H'_{sp}(\brp,\bp',t)$ in terms of the old Hamilton function
$H_{sp}(\br,\bp,t)$ of \eq{htrap}, we used the coordinate transformation of
\eq{trans} and stepped through the Lagrangian procedure.  More elegantly, one
can also employ the  general approach of canonical transformations
\cite{goldstein}.
In particular, we will choose a time-dependent generating function
\begin{align}
\label{generating}
G(\br,\bp'',t)=[\br-\brho(t)]\bp''+\frac{m}{\mathcal{M}}[s(t)+\br
\boldsymbol{\pi}(t)]
\end{align}
that depends explicitly on the old particle coordinate and the new momentum
variable $\{\br, \bp''\}$. It also depends parametrically on the coordinate and
momentum of the drop capsule $\{\brho(t),\boldsymbol{\pi}(t)\}$.  For
convenience, another, yet undetermined variable $s(t)$ was introduced.
This does not affect the dynamics at all, but can be used to match the
energy-zero level at some instant.  The new coordinate $\br''$ and the old
momentum $\bp$ are obtained from the generating function as
\begin{alignat}{2}
  \br''&=\frac{\partial G}{\partial \bp''}=\br-\brho(t), &\quad
  \bp&=\frac{\partial G}{\partial
    \br}=\bp''+\frac{m}{\mathcal{M}}\boldsymbol{\pi}(t).
\end{alignat}
Note that besides the displacement of the coordinate $\br$, we also introduced
a boost in momentum space, which differs from the transformation described by
\eq{trans}.  With respect to the transformed frame, the new Hamiltonian reads
formally as
\begin{equation}
  \begin{aligned}
    \label{h1primeclass}
    H''_{sp}(&\br'',\bp'',t)=
    H_{sp}(\br,\bp,t)+\frac{\partial}{\partial t} G(\br,\bp'',t).
  \end{aligned}  
\end{equation}
By inserting the coordinate transformation as well as employing the Taylor
expansion of the gravitational potential, we find the residual harmonic
Hamiltonian
\begin{equation}
    \label{hdprime}
    H''_{sp}(\br'',\bp'',t)=\frac{{\bp''}^2}{2 m}+ m
    \tilde{V}_t(\br'',\brho(t),t),
\end{equation}
provided the motion of the capsule is externally determined by the
solution of \eq{dyncap} and $s(t,t_0)$ is an action
\begin{equation}
\label{action_s}
s(t,t_0)=\int_{t_0}^{t} \text{d}t'
  L_c[\brho(t'),\dot{\brho}(t'),t']-\brho \pi|^{t}_{t_0}.
\end{equation}
By construction it is clear that we must recover the  identical
\eq{resclasspost} from Hamilton's equations in the new coordinates 
\begin{alignat}{2}
  \dot{\br}''&=\bna{\bp''}{H''_{sp}}, &\qquad
  \dot{\bp}''&=-\bna{\br''}{H''_{sp}}.
\end{alignat}
We would like to point out that these considerations are not purely academic but
will enlighten the analogy between the classical treatment and the quantum mechanical
description of the falling, trapped ultra-cold quantum gas.

\subsubsection{Order of magnitude estimates}  
In order to estimate the magnitude of the Taylor coefficients of the
gravitational potential, one can evaluate them at the surface of the Earth
$\rho=R_\eas=6372$~km and obtains $v_g^{(1)}=9.81\ \mbox{m}\, \mbox{s}^{-2}$,
$v_g^{(2)}=-1.54 \times 10^{-6}\ \mbox{s}^{-2}$ and $v_g^{(3)}=2.42
\times10^{-13}\ \mbox{m}^{-1}\,\mbox{s}^{-2}$.  In there, we made the
approximation of an isotropic gravitational potential.  It is worthwhile
pointing out that $v_g^{(1)}$ is not yet identical to the gravitational
acceleration $g$, which also includes centrifugal corrections (see
Sec.~\ref{rotframe}).


The important consequence of Eq.~(\ref{resclasspost}) is that in first order
$V_g^{(1)}$, the internal harmonic oscillator motion is decoupled from the
center-of-mass motion of the drop capsule.  Only when considering the weak
quadratic correction $V_g^{(2)}$, one can observe a coupling of these motions
as a result of the decrease of the harmonic oscillator frequency along the
gradient of the gravitation.  The following quantum mechanical calculations
will obviously lead to the same conclusions due to Ehrenfest's theorem,
where in case of quadratic Hamiltonians expectation values coincide with the
classical trajectories.  Only by considering the third order correction
$\delta V_g^{(3)}$, we will find an additional dynamical mixing between the
classical trajectory and the quantum mechanical expectation value.

\section{Quantum physics of identical atoms falling within the drop capsule}
\label{secaccel}

So far, we have considered a single classical particle that is harmonically
trapped and falling in the gravitational field. Actually, we are interested in
the behavior of a freely falling, trapped atomic Bose-Einstein condensate
(BEC), which requires a quantum field theoretical description.  However, we do
not want to constrain our discussion solely to condensed bosonic gases, but
would rather like to treat the general situation of an ultra-cold degenerate
quantum gas, whether bosonic or fermionic.

\subsection{Quantized atomic fields}

Before we proceed to the quantum field theoretical description, which is 
intrinsically tied to the picture of second quantization, we briefly revisit
the most prominent relations of first quantization.
In general, the commutator of position and momentum operator $\bro$ and $\bpo$
is
\begin{gather}
  [\bro,\bpo]=i\hbar\, \mathds{1}.
\end{gather}
In the remainder of this article, we will omit the hats for operators in
first quantization, e.g. write $\br$ and $\bp$.
The derived commutation relation for the components of the angular
momentum $\bl=\br\times\bp$ satisfies the standard angular momentum
algebra
\begin{gather}
  \label{angmomentum}
  [l_k,l_l]=i \hbar\,\epsilon_{klm}\, l_m,
\end{gather}
defined through the completely anti-symmetric Levi-Civita tensor
$\epsilon_{klm}$ as structure constants.  Most of the time, we will use the
position representation in this article, hence will interchangeably refer to
$\br$ and $\bp=-i\hbar\bna{\br}{}$ also as the operators of position and
momentum, when acting on Hilbert-space functions.

In order to incorporate the correct quantum statistics according to the Pauli
principle, we use the language of second quantization and introduce the basic
quantities in the following. The field operator in the position representation
will be denoted by $\aop{\br}$ and it annihilates a particle at the position
$\br$, while its hermitian conjugate $\aopd{\br}$ creates it there.  In the
present discussion, all operators are given in the Schr\"odinger picture
unless indicated otherwise. Thus, the quantum field satisfies a commutation
relation for bosons $(-)$ and an anti-commutation rule for fermions $(+)$
\begin{equation}
  \label{aadagger}
  \left[\hat{a}(\br),\hat{a}^{\dagger}(\br') \right]_\mp = 
  \delta\left( \br - \br' \right).
\end{equation}
For commutators, we will drop the ``$-$'' in the following.

Now, we can proceed to single-particle operators in Fock space representing
the center-of-mass coordinate $\Rop$, the linear momentum $\Pop$, as well as
the angular momentum $\Lop$ and the total atom number $\Nop$ as
\begin{equation}
  \label{singparticleops}
  \begin{gathered}
    \Rop=\int\,\text{d}^3 r\,\aopd{\br}\,\br\,\aop{\br},\ \
    \Pop=\int\,\text{d}^3 r\,\aopd{\br}\, \bp\,\aop{\br},\\
    \Lop=\int\,\text{d}^3r
    \,\aopd{\br}\,\bl\,\aop{\br},\ \
    \hat{N}=\int\,\text{d}^3 r\,\aopd{\br}\,\aop{\br}.\\
  \end{gathered}  
\end{equation}
Using the commutation or anti-commutation relation of Eq.~(\ref{aadagger}),
one can easily verify that those operators form a closed algebra
\begin{gather}
  \label{rp}
  \left[\Rop,\Pop\right] = i\hbar\,\Nop,\\
  \label{nparticle}
  \left[\Rop, \Nop \right]=\left[\Pop, \Nop\right] = 
  \left[\Lop, \Nop\right] =0,\\
  \label{pomegal}
  \left[ \Rop, \boldsymbol{\Omega} \Lop \right] 
  =  i \hbar \ \boldsymbol{\Omega} \times \Rop, \quad
  \left[ \Pop, \boldsymbol{\Omega} \Lop \right] 
  =  i \hbar \ \boldsymbol{\Omega} \times \Pop.
\end{gather}
It is indicated by Eq.~(\ref{nparticle}) that this algebra of Euclidean
transformations can be represented in the $N$-particle sector and
$\boldsymbol{\Omega}$ represents an arbitrary rotational axis in
three-dimensional Euclidean space.

In typical BEC experiments, as well as in recent experiments involving
degenerate fermions, the particle densities are of the order $10^{13}$
cm$^{-3}$.  Thus, we can express the energy of a dilute atomic gas of $N$
particles in form of a cluster expansion
\begin{equation}
  \label{HamiltonN}
  \begin{gathered}
    H^{(N)}(t)=\sum_{i=1}^N  H_{{sp}}(\br_i,\bp_i,t)+
    \sum_{i<j}^{N} V_{p}(\br_i-\br_j)
  \end{gathered}
\end{equation}
in terms of single-particle Hamiltonians $H_{sp}(\br,\bp,t)$, pair-wise
interactions $V_{p}(\bxi)$ and, with decreasing relevance, higher order
effects, which can be neglected here \cite{pethick02,stringaribuch}.  
The recently discovered Efimov-resonances \cite{grimmnature} represent 
a very unusual exception to this rule, where genuine three-body effects become important.  
However, the
principle of translational invariance can still be assumed for those
potentials. For the present discussion, we will disregard such peculiarities,
however. Obviously, we want to assume in here that the quantum dynamics of the
$N$-particle state follows the $N$-particle Schr\"odinger equation
\begin{equation}
  \label{SchoedingerN}
  \begin{gathered}
    i\hbar\,\frac{d}{dt}|\Psi^{(N)}\rangle=
    H^{(N)}(t)\,|\Psi^{(N)}\rangle.
  \end{gathered}
\end{equation}

Now, if this is translated to the view point of second quantization, one
obtains the standard Hamilton operator
\begin{equation}
  \label{inertialham}
  \begin{gathered}
    \hat{H}(t)= \hat{H}_{sp}(t)+\hat{V}_{p},\\ 
    \hat{H}_{sp}(t)= \int\,\text{d}^3r\,
    \aopd{\br} H_{{sp}}(\br,\bp,t)
    \aop{\br},\\
    \hat{V}_{p}=\frac{1}{2}\int\text{d}^6 r\ \aopd{\br_1}
    \aopd{\br_2}  V_{p}(\br_1-\br_2) \aop{\br_2} \aop{\br_1}.
  \end{gathered}
\end{equation}
In low-energy field theory, the pair potential $V_{p}$ is often modeled not by
a finite range potential, but it is assumed to be proportional to a delta
function. This is called the contact- or pseudo-potential approximation. It
is very useful in certain situations, but needs careful consideration when
going to higher order calculations \cite{abrikosov65}. For the present
discussion, we can postpone this question and simply demand translational
invariance.

In principle, one can represent any state in Fock space as
\begin{equation}
  \begin{gathered}
    \ket{\Psi}= \sum_{N=0}^{\infty} \int\text{d}^{3N}r\,
    \ket{\br_1,\ldots,\br_N}
    \scp{\br_1,\ldots,\br_N}{\Psi^{(N)}},\\
    \ket{\br_1,\ldots,\br_N}=\aopd{\br_1}\ldots\aopd{\br_N} \ket{v},
  \end{gathered}
\end{equation}
as a superposition of different $N$-particle states $\ket{\br_1,\ldots,\br_N}$
created out of the vacuum $\ket{v}$ and weighted by the properly symmetrized
or anti-symmetrized $N$-particle amplitudes 
$\Psi^{(N)}(\br_1,\ldots,\br_N)=\scp{\br_1,\ldots,\br_N}{\Psi^{(N)}}$.

Proceeding formally, it is also well-known that the complex-valued
Lagrangian for Schr\"odinger fields \cite{hill51,gitman,kobe06}
\begin{gather}
  \mathcal{L}[\Psi(t),\dot{\Psi}(t),t]=\bra{\Psi(t)}
  i\hbar\bna{t}{}-\hat{H}(t)
  \ket{\Psi(t)}
\end{gather}
leads via a Euler-Lagrange variational procedure 
\begin{alignat}{2}
  \Pi(t)&=\frac{\delta \mathcal{L}}{\delta \dot{\Psi}}
  =i\hbar\,{\Psi}(t)^\ast, &\quad \dot{\Pi}(t)&=\frac{\delta
    \mathcal{L}}{\delta \Psi},
\end{alignat}
also to the Schr\"odinger equation in Fock space
\begin{equation}
  \label{schroedinger}
  i\hbar\,\frac{d}{dt}\ket{\Psi(t)}=\hat{H}(t)\,\ket{\Psi(t)},
\end{equation}
finally. With this fundamental equation of motion, we have now established all
the basic concepts and can turn to the transformation of a wave function from
an inertial frame of reference to an accelerated, comoving frame of
reference.

\subsection{Static representation of states in Euclidean frames of reference} 
In this subsection, we will briefly review basic transformation theory for wave functions
under spatial translations and rotations of the coordinate system
\cite{gottfried,schmid77a,schmid77b,klink1,klink2}.

\subsubsection{Three-dimensional Euclidean space}
The basic operation of a translation in space is given by
\begin{equation}
T_{\mathcal{R}}[\br] =\br+\Rc.
\end{equation}
The whole set of translational operators forms an Abelian group \ie
$T_{\mathcal{R}_1}T_{\mathcal{R}_2}=T_{\mathcal{R}_1+\mathcal{R}_2}$. Each
element of this continuous Lie group is indexed by a three-dimensional vector
$\Rc$. 

A rotation around an axis $\boldsymbol{\Omega}$ for a time $t$ is characterized by a
total rotational vector $\Qcb=\boldsymbol{\Omega}\, t$. This operation will be
denoted by $M_{\Qc}$. The set of all those isometric, linear transformations
forms the infinite-dimensional Lie group ${\rm SO}(3)$, which excludes the
possibility of reflections. The entirety of operations establishes the
Euclidean group and a particular element is denoted by
\begin{eqnarray}
  \br'&=&T_{\mathcal{R}} [M_\Qc\,\br]=M_\Qc\,\br+\Rc.
  \label{liegroup}
\end{eqnarray}
If we represent the basic position vector in Cartesian coordinates as 
$\br=\sum_{i=1}^3 r^i
\mathbf{e}_i$, then one finds a three-dimensional orthogonal rotational matrix
\begin{eqnarray}
  M_\Qc\,\br&=&e^{-i/\hbar\, \Qcb \boldsymbol{\ell}}\ \br=
  \br+\Qcb\times\br+ \ldots,
\end{eqnarray}
in terms of the generators of the rotation $\{\ell_1,\ell_2,\ell_3\}$.  For
convenience, we have introduced those $3\times 3$-matrices as complex-valued
angular momenta in units of $\hbar$, which satisfy the angular momentum
algebra of Eq.~(\ref{angmomentum}). No quantum aspects are implied by this
choice of notation (see also Appendix \ref{secb}).

\subsubsection{Single-particle Hilbert space}
The action of the translational operator in single-particle Hilbert space
\begin{gather}
  U_\mathcal{R}=e^{-i/\hbar\,\Rc \bp},
\end{gather}
can be seen most clearly by applying it to position eigenstates
\begin{gather}
  \ket{\br}'=U_\mathcal{R}\,\ket{\br}=\ket{T_{\mathcal{R}}[\br]}=\ket{\br+\Rc}.
\end{gather}
Obviously, the operation 
\begin{gather}
  U_\mathcal{P}=e^{i/\hbar\,\Pc \br},
\end{gather}
is just a boost in momentum space, when acting on momentum eigenstates, \ie
\begin{gather}
  \ket{\bp}'=U_\mathcal{P}\,\ket{\bp}=\ket{T_{\mathcal{P}}[\bp]}=\ket{\bp+\Pc}.
\end{gather}
The unitary representation of the rotation
\begin{gather}
  U_\Qc=e^{-i/\hbar\, \Qcb \bl},
\end{gather}
is induced by the angular momentum operator $\bl$ of \eq{angmomentum}.  When
acting on position as well as momentum eigenstates, one gets
\begin{gather}
  \ket{\br}'=U_\Qc\,\ket{\br}=\ket{M_\Qc\,\br},\\
  \ket{\bp}'=U_\Qc\,\ket{\bp}=\ket{M_\Qc\,\bp}.
\end{gather}
Thus, a representation of a general element of the Euclidean group in Hilbert
space is
\begin{gather}
  \label{unityfirst}
  U=U_\Sc \,U_\mathcal{P}\, U_\mathcal{R}\,  U_\Qc,\quad 
  U_\Sc=e^{i/\hbar\,\Sc}.
\end{gather}
This definition of a general group element is not unique, but admits the
inclusion of an arbitrary phase factor $U_\Sc$, such that the transformed
position state reads
\begin{gather}
  \ket{\br}'=U\,\ket{\br}=
  e^{i/\hbar\,[\Sc+\Pc(M_\Qc\br+\Rc)]}\,
  \ket{M_\Qc \br+\Rc}.
\end{gather}
Such a ray-representation will become physically important when considering
the dynamic evolution of a quantum state, and the accumulated phase will
reflect the  classical action.
\subsubsection{Many-particle Fock space}
We will now proceed to a wave-function $\ket{\Psi}$ in many-particle Fock
space and transform it to the wave-function $\ket{\Psi}'$ in another Euclidean
coordinate system
\begin{equation}
\label{psitrafo}
\ket{\Psi} = \Uop \ket{\Psi}'.
\end{equation}
A general element of the Euclidean group in Fock space 
\begin{gather}
  \label{unitary}
  \Uop=\UN \UP \UR \UO,\\
  \intertext{is now given by the following representation}
  \begin{aligned}
    \UR&=e^{-i/\hbar\,\Rc\,\Pop}, &\qquad
    \UP&=e^{i/\hbar\,\Pc\,\Rop},\\
    \UN&=e^{i/\hbar\,\Sc\,\Nop},&\qquad 
    \UO&=e^{-i/\hbar\, \Qcb\, \Lop},
  \end{aligned}
\end{gather}
where the single-particle operators $\Rop,\Pop,\Lop$ and $\Nop$ have been
introduced in Eq.~(\ref{singparticleops}).

In the following context, we will not need the action of operators $\hat{U}$ on
state vectors $\ket{\Psi}$ in Fock space, but rather on the scalar amplitudes
of the quantum field in the position representation $\aop{\br}$.  This is
straight forward using the commutator relations of Appendix~\ref{secb} in case
of the phase operator $\UN$ and momentum boost operator $\UP$
\begin{gather}
  \UNdag\,\aop{\br} \, \UN= U_\Sc\, \aop{\br},\quad
  \UPdag\,\aop{\br} \, \UP= U_{\mathcal{P}}\, \aop{\br}.  
\end{gather}
For the spatial shift operator $\UR$ as well as the rotational operator $\UO$,
we need to recall that we are dealing with a scalar field. Then one finds that
the unitary operations in single-particle Hilbert space $U_\mathcal{R}$ and
$U_\Qc$ correspond to the inverse operations in Euclidean space
$T_\mathcal{R}^{-1}$ and $M_\Qc^{-1}$, \ie
\begin{gather}
  \URdag\,\aop{\br} \, \UR= U_\mathcal{R}\, \aop{\br}=
  \aop{T_\mathcal{R}^{-1} \br}=\aop{\br-\Rc},\\
  \UOdag\,\aop{\br} \, \UO= U_{\Qc}^{\vphantom \dag} \, \aop{\br}=
  \aop{M_\Qc^{-1} \,\br}.
\end{gather}

\subsection{Dynamics in comoving frames of reference}
\label{secdyn}
\subsubsection{Frame transformation from the inertial to the comoving frame}
The Euclidean coordinate transformations of the previous section where
determined by ten static parameters, $\{\Sc, \Rc,\Pc,\Qcb\}$.
By considering an arbitrary time-dependence of these parameters, we can
introduce a canonical frame transformation that includes the Galilean
transformation as a special case \cite{gottfried}. Currently, we will focus on
non-rotating frames ($\Qcb=0$) for simplicity, \ie
\begin{gather}
\label{norot}
  \Uop(t)=\hat{U}_{\mathcal{S}(t)} \,
  \hat{U}_{\mathcal{P}(t)} \,
  \hat{U}_{\mathcal{R}(t)} 
\end{gather}
and lift this restriction in Sec.~\ref{rotframe}.

Our goal is to determine an equation of motion for this comoving, accelerated
frame such that the residual motion of the atomic gas within the frame is free
of any non-inertial forces.  Given the Schr\"odinger
equation~(\ref{schroedinger}) holds in the inertial Earth-centered frame, then
we find another realization $\ket{\Psi}=\Uop(t) \ket{\Psi(t)}'$ in an
accelerated frame of reference
\begin{gather}
  \label{schroedingerneu}
  i\hbar\,\frac{d}{dt}\ket{\Psi(t)}'= \hat{H}'(t)\ket{\Psi(t)}',\\
  \intertext{with the transformed Hamiltonian}  
  \label{hop3}
  \hat{H}'(t)=\Uopdag(t)\,(-i\hbar \frac{d}{dt} +\hat{H}_{sp}(t)
  +\hat{V}_{p})\,\Uop(t).
\end{gather}
In here, the first contribution to the Hamiltonian is a gauge potential, in
analogy to the fictitious forces that would appear in the classical Hamiltonian
function in terms of the time derivative of the generating function,
cf. Eq.~(\ref{h1primeclass}). It can be evaluated using
Eqs.~(\ref{displace}, \ref{rotate}) and it generates additional gauge forces
\begin{equation}
  \Uopdag(t)( -i\hbar \frac{d}{dt}) \Uop(t) = (\dot{\Sc}+\dot{\Pc}\Rc)\,\Nop
   +\dot{\Pc}\,\Rop -\dot{\Rc}\,\Pop.
  \label{nb3}
\end{equation}
The second contribution to the Hamiltonian implements the spatial translation
and momentum boost
\begin{gather}
  \Uopdag(t) \, \hat{H}_{sp}(t)\,\Uop(t)\\
  =\int\,\text{d}^3r'\, \aopd{\br'}\,H_{{sp}}(\br'+\Rc,\bp'+\Pc,t)
  \,\aop{\br'}, \nonumber
  \label{nb1}
\end{gather}
where it is implicitly assumed that the one-particle Hamiltonian
$H_{sp}(\br,\bp,t)$ can be expanded into a power series in $\br$ and
$\mathbf{p}$, respectively.

The third term of the Hamiltonian Eq.~(\ref{hop3}) is not affected by the
frame transformation,
\begin{gather}
  \label{nb2}
  \Uopdag \,\hat{V}_{p}\,\Uop=\hat{V}_{p},
\end{gather}
as the interaction of Eq.~(\ref{inertialham}) is hermitian (number
conserving), a local operator, and translational invariant by construction in
Fock-,  as well as in two-particle Hilbert space,
\begin{equation}
  [\Nop,\hat{V}_p]=[\Rop,\hat{V}_p]= [\Pop,\hat{V}_p]=0.
\end{equation}
This property is also the basis of Kohn's theorem \cite{kohn}, \ie the exact
decoupling of the center-of-mass motion and the relative excitations in
maximally quadratic single-particle Hamiltonians. It is also known as the
harmonic potential theorem \cite{dobson} and has been used in the context of cold
dilute quantum gases \cite{birula02, song05}.

Combining these results, we find the transformed Hamiltonian in the
accelerated frame
\begin{equation}
\label{acceham}
\hat{H}'(t)= \int\,\text{d}^3r'\,
\aopd{\br'}\, H_{sp}'(\br',\bp',t)\, \aop{\br'}+\hat{V}_{p}
\end{equation}
and all gauge contributions are contained in the single-particle Hamiltonian
\begin{equation}
  \label{transH}
  \begin{split}
    H_{sp}'(\br',\bp',t)=&H_{sp}(\br'+\Rc,\bp'+\Pc,t)\\
    &+\dot{\Sc} + \dot{\Pc}\,\Rc +\dot{\Pc}\,\br' - \dot{\Rc}\,\bp'.
  \end{split}
\end{equation}

\subsubsection{Application of the frame transformation 
in the harmonic approximation}

We will now apply the preceeding considerations to the falling trapped
interacting many-particle system in the drop tower. The single-particle
Hamilton operator follows from the classical considerations of
Eq.~(\ref{htrap}) and is given by
\begin{equation}
  \label{h1_spec}
  H_{sp}(\br,\bp,t)=\frac{\bp^2}{2\,m}+
  m\,[V_{{t}}\left(\br-\brho(t),t\right)+V_g(\br,t)].
\end{equation}
By splitting the transformed Hamiltonian Eq.~(\ref{transH}) into constant,
linear and higher order polynomial contributions in terms of the position
$\br'$ and the momentum $\bp'$, one finds
\begin{equation}
  \begin{gathered} 
    H_{sp}'(\br',\bp',t)=
    (\boldsymbol{\alpha}\bp'+\boldsymbol{\beta}\br'+\gamma)
    +\frac{\bp'^2}{2m}+mV_{{t}}\left(\br',t\right)\\
    +m\left[V_g(\br'+\Rc,t)-V_g(\Rc,t)-V_g^{(1)}(\br',\Rc,t)\right]
\end{gathered}
\end{equation}
with coefficients 
\begin{equation}
  \label{constraint}
  \begin{aligned}
    \boldsymbol{\alpha}=&\frac{\Pc}{m}-\dot{\Rc},\quad
    \boldsymbol{\beta}=\dot{\Pc}+
    m\,\bna{\Rc}{}[V_{t}(\Rc-\brho,t)+V_g(\Rc,t)],\\
    \gamma=&
    \frac{\Pc^2}{2\,m}+m\,\left[V_{t}(\Rc-\brho,t)+V_g(\Rc,t)\right]
    +\dot{\Sc}+\dot{\Pc}\,\Rc.
  \end{aligned}
\end{equation}  
If the constraints are satisfied identically, \ie
$\{\boldsymbol{\alpha},\boldsymbol{\beta},\gamma\} \rightarrow 0$, one obtains
the single-particle Hamiltonian, if we disregard again terms of third and higher
orders $\delta V_g^{(3)}$ in the accelerated frame,
\begin{gather}
  \label{hprime}
  \begin{aligned}
    H_{sp}'(\br',\bp',t)= &\frac{\bp'^2}{2\,m}
    +m\,\widetilde{V}_{{t}}\left(\br',\Rc(t),t\right),
  \end{aligned} 
 \end{gather}
 provided the frame parameters as well as the center-of-mass of the drop
 capsule  $\{\Rc(t),\Pc(t),\brho(t)\}$ satisfy the classical
 equations of motion 
\begin{alignat}{2}
\label{piandpidot}
\pi&=\bna{\dot{\brho}}{L_c}=\mathcal{M} \dot{\brho},&\quad
\dot{\pi}&=\bna{\brho}{L_c},\\
\label{cmosc}
\Pc&=\bna{\dot{\Rc}}{L_{sp}}=m\dot{\Rc},&\quad
\dot{\Pc}&=\bna{\Rc}{L_{sp}}.
\end{alignat}
We can express the latter line as Newton's equation of motion for $\Rc$,
\begin{equation}
\label{newtonrcal}
\ddot{\Rc}=-\bna{\Rc}{}[V_{t}(\Rc-\brho,t)+V_g(\Rc,t)].
\end{equation}
With Eqs.~(\ref{piandpidot}) and (\ref{cmosc}), we have recovered the Lagrangian $L[\brho,\dot{\brho},\Rc,\dot{\Rc},t]$
of \eq{lagcappluspart}, and one can express the accrued dynamical phase as an
action integral
\begin{gather}
\Sc(t,t_0)=\int_{t_0}^t dt' L_{sp}[\Rc,\dot{\Rc},t']-\Pc\Rc|^t_{t_0},
\end{gather}
like in \eq{action}, where we started from, cf. also Eq.~(\ref{action_s}).

The harmonic trapping potential $\widetilde{V}_{{t}}$ that appears only in the
residual Hamiltonian, \eq{hprime}, has been modified by quadratic corrections of
gravity and was defined in \eq{eq:trappotmod}. The third order correction to
the gravitational potential $\delta V_g^{(3)}$ is minuscule. On the time scale of the
free fall experiment, one may safely disregard it.  Thus, we are left with a
quadratic, time-dependent Hamilton operator and the harmonic potential theorem
applies \cite{dobson}.

As before, the motion of the center-of-mass coordinate $\Rc$,
\eq{newtonrcal}, can be understood much more clearly, if we expand the 
gravitational potential
around the center-of-mass of the drop capsule $\brho$.  For the deviation
$\delta\Rc(t)=\Rc(t)-\brho(t)$ after the release of the capsule,
one finds a pure harmonic oscillation
\begin{equation}
\begin{aligned}
  \label{classR}
  \delta\ddot{\Rc}(t)&=-\tilde{v}^{(2)}(\brho(t),t)\, \delta \Rc(t).
\end{aligned}  
\end{equation}

\subsection{Ehrenfest's theorem}

At this point, we would like to confirm the physical interpretation of the
frame parameters $\Rc(t)$ and $\Pc(t)$ as center-of-mass and momentum
coordinates, respectively. Clearly, they have been introduced as such 
with dimensions of length and momentum. Using Ehrenfest's
theorem, we are able to verify this.

The temporal evolution of the expectation value of any time-independent
operator $\hat{A}$, \ie $\ave{\hat{A}}=\bra{\Psi(t)} \hat{A} \ket{\Psi(t)}$
follows from the Schr\"odinger \eq{schroedinger} and gives
\begin{equation}
  i\hbar\,\frac{d}{dt} \langle \hat{A} \rangle =
  \langle [\hat{A},\hat{H}(t)]\rangle.
\end{equation}
By using the bosonic or fermionic commutation relation, Eq.~(\ref{aadagger}),
we can work out the commutators in the Schr\"odinger picture 
\begin{align}
  \frac{i}{\hbar}[\hat{H}(t),  \Rop ]=&\frac{\Pop}{m},\\
  \frac{i}{\hbar}[\hat{H}(t), \Pop ]=& -v^{(1)}_g(\brho(t))\,\Nop\\
  &-\tilde{v}^{(2)}(\brho(t),t)\,(\Rop-\brho(t)\Nop), \nonumber
\end{align}
where we have discarded third order corrections. In analogy to \eq{classR}, we
can define a deviation from the drop capsule coordinate as $\delta \langle
\Rop \rangle=\langle\Rop\rangle - \brho(t)N$ and find that it evolves
according to
\begin{gather}
  \label{res_ehren}
  \delta \langle \ddot{\hat{{\bf{R}}}} \rangle= -\tilde{v}^{(2)}(\brho(t),t)
  \,\delta \langle \Rop \rangle,\qquad
  \ave{\Pop}=m\,\langle\dot{\hat{{\bf{R}}}} \rangle,
\end{gather}
where we have introduced the particle number $N=\langle \Nop \rangle$.
Indeed, Eqs.~(\ref{resclasspost}), (\ref{classR}) and (\ref{res_ehren}) are
identical in form, which is just Ehrenfest's theorem.  Furthermore, if we
assign the identical initial conditions to the frame parameters
$\Rc(t_0)=\langle\Rop(t_0)\rangle/N$ and $\Pc(t_0)=\langle \Pop(t_0) \rangle/N$,
then we can identify them with the center-of-mass coordinate and total
momentum of the atomic ensemble.

\subsection{Application of the comoving frame transformation 
  to mean-field theory}
All the calculations that were presented so far are generally valid on the
many-particle level using the language of second quantization.  We have seen
that the underlying physics of frame transformations in translational
invariant systems rests exclusively on the single-particle level. If we
furthermore restrict our scope to harmonic potentials, everything can be
understood in terms of classical physics, in principle.

\subsubsection{From quantum fields to classical fields} 

For the practical purpose of studying the internal dynamics of Bose-Einstein
condensates or superfluid fermionic gases in $\mu$-gravity, it is usually
sufficient to restrict oneself to the mean-field picture. However, the
application of the presented frame transformations to an extended mean-field
approach in the case of bosons \cite{walser599} or fermions
\cite{kokkelmans502} is straight forward.

In the case of a bosonic gas, the Gross-Pitaevskii (GP) equation provides a
very successful description of many dynamical and static properties of the
BEC \cite{dalfovo}. To derive the GP equation in the accelerated frame, we
start from Heisenberg's equation for an operator $\hat{A}$, which is
time-independent in the Schr\"odinger picture
\begin{equation}
  i\,\hbar\,\frac{d}{dt}\hat{A}_H(\br,t)=
  [\hat{A}_H(\br,t),\hat{H}'_H(t)],
\end{equation}
which can be obtained from the Schr\"odinger equation in the accelerated frame
\eq{schroedingerneu} and the Hamilton operator \eq{acceham}.  The
representation of operators in Heisenberg picture requires the introduction of
the time-ordered ($\mathcal{T}$) evolution operator
\begin{gather}
  \hat{A}_H(\br,t)=\hat{T}^{\dagger}(t,0)\,\hat{A}(\br)\,\hat{T}(t,0),\\
  \hat{T}(t,0)=\mathcal{T}\exp\left[-i/\hbar
    \int_{0}^{t}\text{d}\tau\,\hat{H}'(\tau) \right].
\end{gather}
In particular, one finds for the field operator
\begin{gather}
  \begin{gathered}
    \label{heisa}
    i \hbar\,\frac{d}{dt}\hat{a}_H(\br',t)=
    \left[H_{sp}'(\br',\bp',t)\right.\\
    \left.+\int\text{d}^3\,\xi\
      \hat{a}_H^{\dagger}(\bxi,t)\,
      V_{p}(\bxi-\br')\,\hat{a}_H(\bxi,t)\right]
    \hat{a}_H(\br',t).
  \end{gathered}
\end{gather}
In the case of a dilute weakly correlated BEC, it is usually sufficient to
stay within the mean-field approximation. This is the classical limit and
replaces the quantum field $\hat{a}_H(\bxi,t)$ by a complex valued field
$\alpha(\bxi,t)$. This approximation is also consistent with the use of a
pseudo potential for the pair interaction \ie
\begin{equation}
  V_p(\bxi)=\frac{4\pi \hbar^2 a_s}{m}\,\delta(\bxi)=\kappa\,\delta(\bxi).
\end{equation}
It uses the s-wave scattering length $a_s$ in vacuo and basically summarizes
all the binary scattering contributions to the Born series that are still
present in the full Heisenberg \eq{heisa}.  With these assumptions, one
arrives at the GP equation in the accelerated frame
\begin{equation}
  i\hbar\,\partial_t\alpha(\br',t)=
  \left[H_{sp}'(\br',\bp',t)
    +\kappa\,|\alpha(\br',t)|^2 \right]\,\alpha(\br',t),
\end{equation}
with the Hamiltonian of \eq{hprime} and the classical frame equations 
\eq{cmosc}.

\subsubsection{Numerical study of the long time evolution of a 
  freely falling condensate} 
\begin{figure}[h]
\begin{center}
\vspace{0.1cm}
\includegraphics[width=\columnwidth,angle=0]{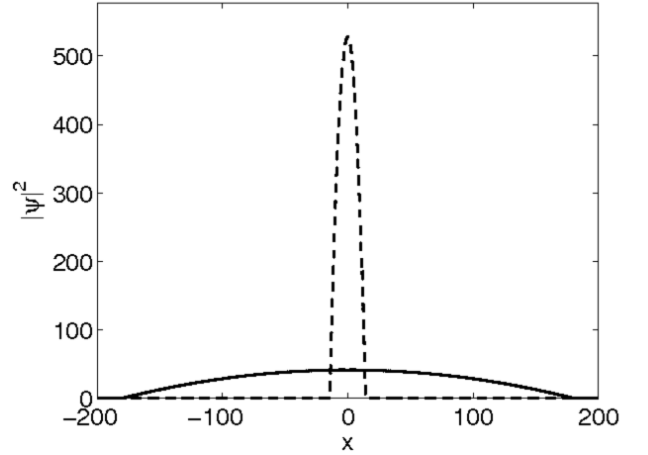}
\caption{\label{freeprop} 
  Free expansion of an initially trapped BEC over 5 seconds.  Length is
  measured in harmonic oscillator units. The dashed line shows the
  density of the BEC at the instant of release, the solid line shows the
  condensate at the end of the free expansion.}
\end{center}
\end{figure}

\begin{figure}[h]
\begin{center}
  \vspace{0.1cm} \includegraphics[width=\columnwidth,angle=0]{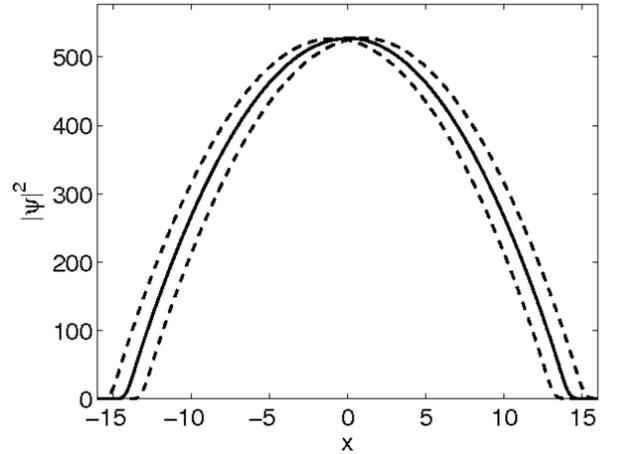}
\caption{\label{pendulum} 
  Oscillatory motion of a condensate that is initially displaced by $x_0=1\,
  a_{ho}$ from the equilibrium position. The Kohn mode oscillates with the
  trap angular frequency $\omega_z$ and decouples from the free fall dynamics.
  The solid line shows the density of the BEC at equilibrium, the dashed lines
  correspond to the turning points.}
\end{center}
\end{figure}

In order to illustrate the benefit of the
transformation to a comoving frame, we present a simple numerical calculation of a BEC
in a time-dependent trap after this transformation. Within the GP mean-field
picture, we calculated the corresponding time-evolution for two time-dependent
trap configurations. For simplicity, we considered a quasi one-dimensional BEC
of $^{87}$Rb atoms and solved the corresponding GP equation. This is possible
for a very prolate trap configuration, $\gamma(t)=\omega_r/\omega_z(t)\gg 1$,
with the radial and longitudinal oscillator angular frequency $\omega_r$ and
$\omega_z(t)$, respectively. The radial motion is effectively frozen out.
Then, we can separate the ground state of the harmonic oscillator in radial
direction and integrate out this degree of freedom.  This yields
\begin{equation}
i\hbar\,\partial_{t}\alpha=\left(-\frac{\hbar^2}{2m}\partial_{z}^2
+\frac{m\omega_z(t)^2}{2}z^2
+\kappa'|\alpha|^2\right)\alpha.
\end{equation}
In there, $\alpha=\alpha(z,t)$ is the order parameter within the
one-dimensional GP equation in the comoving frame of reference. A constant
energy-offset arising from the integration over the radial part has been
removed by switching to an interaction picture. $\kappa'$ is the coupling
constant modified due to the radial integration.  In Figs.~\ref{freeprop} and
\ref{pendulum}, we show the time evolution of a $^{87}$Rb condensate
consisting of $N=10000$ atoms for two situations: In the first case, the free
time evolution is studied, i.e. the trap is switched off instantaneously at
$t=0$, when starting the drop experiment, which corresponds to
$\omega_z(t)=\omega_z\,\Theta(-t)$ with the well-known Heaviside function. Due
to the duration of about 5 seconds, the BEC can expand largely. Our chosen set
of parameters is the atomic mass of $^{87}$Rb $m=86.9$ amu (unified atomic
mass units), $\omega_z=1$/s, and $\omega_r=2\pi\,100$/s.  In the figures, we
used scaled quantities, i.e.  time is normalized to the longitudinal
oscillator frequency, $\bar{t}=t\,\omega_z/2\pi$, lengths are measured in
harmonic oscillator units $a_{ho}=\sqrt{\hbar/m\,\omega_z}= 27 \mu\text{m}$.
The scaled coupling constant is given by
$\bar{\kappa}'=2\gamma(0)\,a_{s}/a_{ho}$, where the s-wave scattering length
$a_{s}=5.8$ nm. For the $^{87}$Rb parameters see for example
\cite{dshall98}.  In the second case, we simulated the time
evolution of a condensate, which is displaced from its equilibrium in the trap
by one scaled length unit at the instant of release.  The oscillator motion
decouples from the free fall trajectory. In here, we assumed a constant trap
angular frequency $\omega_z=1$/s.

\section{\label{rotframe} Many identical atoms within the drop capsule
  in a rotating frame}

In the previous sections, we have deliberately omitted the rotation of the
Earth from our discussion in order to simplify the algebra and focus on the
essential physics. In particular, we only considered the transformation
from the Earth-fixed inertial frame $\{\mathbf{e}_i\}$, located at its origin, 
to the accelerated frame of the freely falling, center-of-mass
coordinate $\Rc(t)$ of the atomic cloud (see Fig.~\ref{coord1}).

\subsection{Classical physics of the drop capsule in a rotating 
frame of the Earth}

\subsubsection{Classification of the three important frames of reference}

\begin{figure}[h]
  \begin{center}
    \includegraphics[width=\columnwidth]{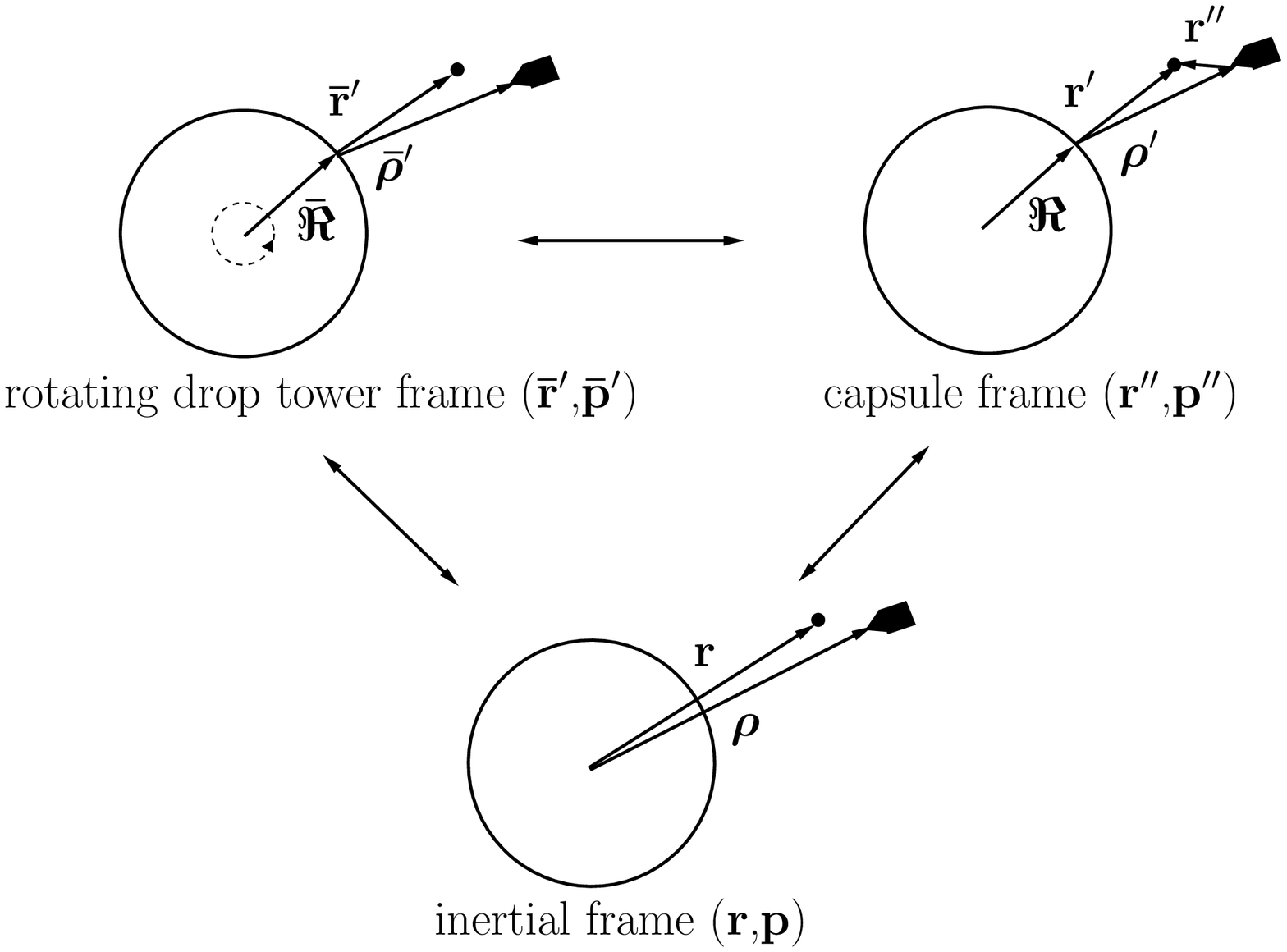}
    \caption{\label{systems} 
      Schematic view of the three different reference frames as seen from
      above the North pole. In the inertial frame, located at the center of
      the Earth, a particle is characterized by $(\mathbf{r},\mathbf{p})$, and
      the drop capsule is at position $\brho$.  In the corotating drop tower
      frame, with the origin at the bottom of the tower, these quantities are
      denoted by $(\brbp,\bpbp)$ and $\brhob'$, while in the drop capsule
      frame the particle's coordinate and momentum are
      $(\mathbf{r}'',\mathbf{p}'')$.}
  \end{center}
\end{figure}

In fact, in our considerations three different frames of reference are of
significant importance. Besides the inertial frame,
a real experiment obviously requires a description in a coordinate
system that is aligned with the corotating drop tower tube. 
Moreover, in the frame comoving with the drop capsule,
the experiment can be described in a most simple way. In this frame of reference,
non-inertial forces can be eliminated. It is of special importance to describe the
mapping between the drop tower frame and the capsule frame, since both systems
can be used for observation, in principle.
In the inertial frame, a point-like particle is characterized by 
$(\mathbf{r},\mathbf{p})$, and the drop capsule is at position $\brho$.
In the corotating drop tower frame, with origin at the bottom of the 
tower, we indicate these quantities with $(\brbp,\bpbp)$ and $\brhob'$,
while in the drop capsule frame the particle's coordinate and momentum are $(\mathbf{r}'',\mathbf{p}'')$. A schematic sketch of the three systems is 
depicted in Fig~\ref{systems}.


\subsubsection{Characterization of the drop tower frame}

\begin{figure}[h]
  \begin{center}
    \includegraphics[width=\columnwidth]{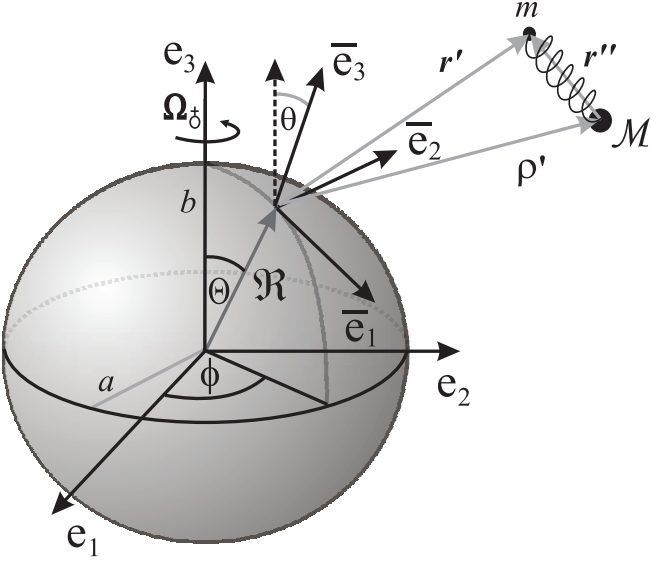}
    \caption{\label{coord} A harmonically trapped particle with position
      $\brp$ and mass $m$ in the drop capsule with position $\brho'$ and
      mass $\mathcal{M}$. The position in the capsule
      frame is denoted by $\br''$.  The drop tower axes
      $\{\mathbf{\bar{e}}_i\}$ perform a diurnal rotation around the Earth
      axes $\Oeas$. $\theta$ denotes the angle between the rotational axis of
      the Earth and the plummet at the geographic location $(\phi,\Theta)$ of
      the drop tower $\Reas$. The oblate ellipsoidal shape of the Earth is
      modeled by the half axes $a$ and $b$.}
  \end{center}
\end{figure}

In the rest frame of the drop tower, the coordinate
system is aligned with the rotating tower principal axes denoted by
$\{\mathbf{\bar{e}}_j(t)\}$. This is depicted in Fig.~\ref{coord}. 
$\theta$ denotes the angle between the rotational axis of 
the Earth and the plummet at the geographic location $(\phi,\Theta)$ of the 
drop tower $\Reas$. Due to the ellipsoidal shape of the Earth, the plummet 
does not point to the center of the Earth. The oblateness is modeled by the 
half axes $a$ and $b$.
The instantaneous principal tower axes, as well as the direction towards the
base of the tower $\Reas(t)$
\begin{alignat}{2}
  \label{M}
  \mathbf{\bar{e}}_i(t)&=M_{\small \eas}(t)\,
  M_{\phi \mathbf{e}_3}\,M_{\theta \mathbf{e}_2} \,\mathbf{e}_i,
  &\quad 
  \dot{{\bar{{\bf e}}}}_i&=\Oeas\times \mathbf{\bar{e}}_i,\\
  \Reas(t)&=M_{\small \eas}(t)\,M_{\phi \mathbf{e}_3}\,M_{\Theta \mathbf{e}_2}\,
  \mathbf{e}_3,
  &\quad
  \dot{\Reas}&=\Oeas\times \Reas,
\end{alignat}
are obtained from the inertial axes of the Earth $\{\mathbf{e}_i\}$ by
aligning the $\mathbf{\bar{e}}_3$-axis of the tower along the direction of the
plummet, followed by a rotation $\phi$ to the longitude of the tower
at ZARM/Bremen, as well as the diurnal rotation $M_{\small \eas}(t)$ around
the Earth axis $\Oeas=\Oe\, \mathbf{e}_3$ with an angular frequency
$\Oe=7.2\times10^{-5}$~1/s. No further time-dependence caused by geophysical
effects, like tidal forces, precession, or wobbling due to the liquid core
motion will be considered.

Now we are in the position to represent all vectors either in the inertial
Cartesian basis of the Earth or the corotating frame \ie
\begin{align}
  \mathbf{q}(t)&=\sum_{i=1}^{3}q^i(t)\,\,\mathbf{e}_i=
  \sum_{j=1}^{3}\bar{q}^j(t)\,\,\mathbf{\bar{e}}_j(t),\\
  q_i&=\sum_{j=1}^{3}\mathbf{e}_i \mathbf{\bar{e}}_{j}\,\bar{q}^j=\sum_{j=1}^{3}M_{ij}(t)\,\bar{q}^j,
\end{align}
by introducing the matrix representation of the orthogonal 
rotation  $M(t)^\top M(t)=\mathds{1}$ defined by
\eq{M}.  It is obvious that the components of the Earth rotation as well as
the pointer to the base of the tower
\begin{align}
  \bar{\Omega}_{\tiny \earth}^i&= \Oe \,(-\sin\theta,0,\cos \theta ),\\
  \bar{\Re}^i&= \Re \,(\sin{(\Theta-\theta)},0,\cos{(\Theta- \theta)} ),
\end{align}
must be time independent in this basis.

\subsubsection{Dynamics of the drop capsule}
All distances will be measured relative to the base of the drop tower
$\Reas(t)$, \ie
\begin{align}
  \brho=\brho'+\Reas=M(t)\,\bar{\brho}=M(t)\,(\brhob'+\bar{\Reas}).
\end{align}
By differentiation, one finds the relation between the velocities with respect
to the two frames as
\begin{align}
\label{rot}
  \dot{\brho}&=M(t)\,(\dot{\brhob}'+ \Oeasb\times \brhob).
\end{align}
In the rotating frame, the Lagrangian of \eq{lag1} for the relative
position $\brhob'$ as well as the relative velocity
$\dot{\brhob}'$ reads
\begin{equation}
\label{Lcbar}
  \begin{gathered}
    \bar{L}_c'[\brhob',\dot{\brhob}',t]
    =\mathcal{M}\left(\frac{(\dot{\brhob}'+\Oeasb\times\brhob)^2}{2} 
      -V_g(M(t) \brhob,t)\right)\\
    =\mathcal{M}\left(\frac{\dot{\brhob}'^2}{2} 
      -V_c(\brhob)-V_g(M(t) \brhob,t) \right)
+\Oeasb  (\brhob\times \dot{\brhob}' \mathcal{M} ).
  \end{gathered}  
\end{equation}
Here, we have introduced the centrifugal potential as
\begin{equation}
  \label{potcent}
  \begin{aligned}
    \bar{v}_c^{(2)}& =\Oeasb\otimes\Oeasb-{\Oe^{2}},\\  
    V_c(\brhob)&=\frac{1}{2} (\brhob\otimes \brhob)
    \cdot \bar{v}_c^{(2)}.
  \end{aligned}
\end{equation}

The canonical momentum $\bpib'$, that is conjugate to
the coordinate $\brhob'$, is given by
\begin{align}
  \bpib'=\bna{ \dot{\brhob}'}{\bar{L}_c'}= \mathcal{M}
  (\dot{\brhob}'+\Oeasb\times\brhob)
\end{align}
and implies a Hamiltonian in the new coordinates
\begin{equation}
\begin{split}
  \bar{H}_c'(&\brhob',\bpib')
  =\bpib' \dot{\brhob}'-\bar{L}_c'\\
  &=\frac{\bpib'^2}{2 \mathcal{M}}+
  \mathcal{M} V_g(M(t) \brhob,t)- \Oeasb\bar{\bf{l}}_c',
\end{split}
\end{equation}
where the angular momentum 
$\bar{\bf{l}}_c'= \brhob\times \bpib'$ has been
introduced.
Via Hamilton's equations 
\begin{align}
  \label{hamroteq_1} 
  \dot{\brhob}'&=\bna{\bpib'}{\bar{H}_c'}=\frac{\bpib'}{\mathcal{M}}
  -\Oeasb\times \brhob,\\
\label{hamroteq_2} 
\dot{\bpib}'&=-\bna{\brhob'}{\bar{H}_c'}= -\mathcal{M}\bna{\brhob'}{} V_g(M(t)
\brhob,t) -\Oeasb\times \bpib',
\end{align}
we are lead directly to the well-known 
Newton equations  in the rotating frame \cite{greiner}
\begin{gather}
  \label{classicnew}
  \ddot{\bar{\brho}}' 
  = -2 \, \Oeasb \times \dot{\brhob}'
  -\bna{\brhob'}{} [V_c(\brhob)+ V_g(M(t) \brhob,t)],
\end{gather}
where the first term is the Coriolis force, followed by the centrifugal force
of \eq{potcent} and the gravitational force as defined before in \eq{force}.
By expanding the gravitational potential to second order around the base of
the drop tower $\Reas$ (see Fig.~\ref{coord}), we obtain
\begin{align}
\label{classicnewlin}
  \ddot{\bar{\brho}}'=-2 \, &\Oeasb \times \dot{\bar{\brho}}'
  -\bar{\mathbf{g}}(\Reas)
  -\left(\bar{v}^{(2)}_g(\Reas)+\bar{v}_c^{(2)}\right)\,
  \bar{\brho}'.
\end{align}
Here, we have introduced the effective gravitational acceleration and rotated
Taylor coefficients
\begin{align}
  \bar{\mathbf{g}}(\Reas)&=\bar{\mathbf{v}}^{(1)}_g(\Reas)+\Oeasb \times
  \left(\Oeasb \times \bar{\Reas} \right),\\
  \bar{\mathbf{v}}^{(1)}_g(\Reas)&=M(t)^{\top}  \mathbf{v}^{(1)}_g(\Reas),\\
  \bar{v}^{(2)}_g(\Reas)&=M(t)^{\top}v^{(2)}_g(\Reas) M(t),
\end{align}
at the surface of the Earth. $\bar{\mathbf{g}}$ is normal to the surface and
takes into account the ellipsoidal shape of the Earth.  In order to obtain a
simple solution of Eq.~(\ref{classicnewlin}), one can neglect the centrifugal
term, since it is of the order $\Oe^2 \approx 10^{-9}\text{m}/s^2$, see
Appendix C.  An estimate of the Coriolis deviation in
$\bar{\mathbf{e}}_2$-direction after a drop distance of 100 m yields
approximately 2\,cm, which is not negligible.

\subsection{A single classical particle in the drop capsule}

\subsubsection{Transformation from inertial- to tower coordinates}

Now, let us get back to the harmonically trapped particle at $\br$ inside the
drop capsule $\brho$ and consider its motion with respect to the rotating base
of the tower at $\Reas$ analogous to Eq.~(\ref{trans}).
In this case, coordinates and velocities are given by
\begin{align}
  \br&=M(t) \brb=
  M(t) (\brbp+\bar{\Reas}),\\
  \dot{\br}&=M(t)\,(\dbrbp+\Oeasb \times\brb).
\end{align}
Note, that this description refers to the tower coordinates.

The trap rests in the drop capsule, which is moving on the classical
trajectory given by \eq{classicnew}. From the Lagrangian in the
inertial frame, \eq{lagtrap}, we obtain the Lagrangian for the particle
coordinates relative to the base of the drop tower
\begin{equation}
\label{eulapart}
\begin{aligned}
  \bar{L}'_{sp}[&\brbp,\dbrbp,t]
  =m\Big[\frac{1}{2}(\dbrbp+\Oeasb\times\brb)^2\\    &-V_t(M(t)(\brbp-\brhob'),t)-V_g(M(t)(\brbp+\Reasb),t)\Big]. 
\end{aligned}
\end{equation}
The total Lagrangian, which includes the Lagrangians both of the drop capsule as well as the single trapped particle, reads
\begin{equation}
\label{lbarp}
\bar{L}'=\bar{L}_c'+\bar{L}_{sp}'
\end{equation}
with $\bar{L}_c'$ as defined in Eq.~(\ref{Lcbar}).
The linear canonical momentum is
\begin{align}
  \bpbp&=\bna{\dbrbp}{\bar{L}_{sp}}= m \,
  (\dbrbp+\Oeasb\times\brb),
\end{align}
and thus one finds for the Hamiltonian
\begin{align}
\label{Hbarsp}
\bar{H}_{sp}'(&\brbp,\bpbp,t)=
\frac{\bpbp^2}{2 m}+V_t(M(t)(\brbp-\brhob'),t)\nonumber \\
&+V_g(M(t)(\brbp+\Reasb),t)-\bpbp\left(\Oeasb\times\brb\right).
\end{align}
Then Hamilton's equations of motion read
\begin{gather}
\dot{\bar{\mathbf{r}}}'=\frac{\partial\bar{H}_{sp}'}{\partial\bpbp}=
\frac{\bpbp}{m}-\Oeasb\times\brb,\\
\dot{\bar{\mathbf{p}}}'=-\frac{\partial\bar{H}_{sp}'}{\partial\brbp}=
-\frac{\partial}{\partial\brbp}\left(V_t+V_g\right)
+\bpbp\times\Oeasb,
\end{gather}
correspondingly.
Finally, Newton's equation is given by
\begin{equation}
\begin{split}
\ddot{\bar{\mathbf{r}}}'=&-\frac{\partial}{\partial\brbp}\left(V_t(M(t)(\brbp-\brhob'),t)+V_g(M(t)(\brbp+\Reasb),t)\right)\\
&-\Oeasb\times(\Oeasb\times\brb)-2\Oeasb\times\dot{\bar{\mathbf{r}}}'.
\end{split}
\end{equation}
If we expand the gravitational potential up to second order, we obtain
\begin{align}
  \ddot{\brb}'= &-2 \, \Oeasb \times \dbrbp -\bar{\mathbf{g}}(\brhob)
  -\bna{\brbp}{}
  \Check{\Bar{V}}_t(\brbp-\brhob',\brhob,t) \nonumber\\
  =& -2 \, \Oeasb \times \dbrbp
  -\Check{\Bar{v}}^{(2)}(\brhob,t)\,(\brbp-\brbp_{eq}(t)).
\end{align}
As in the non-rotating situation of Sec.~\ref{pre}, we get an equilibrium position
$\brbp_{eq}(t)$, which is defined by
\begin{equation}
  \Check{\Bar{v}}^{(2)}(\brhob,t)\,
  \brbp_{eq}(t)= -\bar{\mathbf{g}}(\brhob)+\Check{\Bar{v}}^{(2)}(\brhob,t)\brhob'
\end{equation}
and a modified, appropriately rotated harmonic trapping potential
\begin{gather}
  \Check{\Bar{v}}^{(2)}(\brhob,t)=
  \bar{v}_t^{(2)}(t)+\bar{v}^{(2)}_g(\brhob,t)+\bar{v}_c^{(2)},\\
  \Bar{v}^{(2)}_t(t)=M(t)^{\top}\, v^{(2)}_t(t) \,M(t),\\
  \Check{\Bar{V}}_t(\bxi,\brhob,t)=\frac{1}{2}\,(\bxi \otimes \bxi)\cdot
  \Check{\Bar{v}}^{(2)}(\brhob,t).
\end{gather}
The equilibrium position $\brbp_{eq}(t)$ is governed by the trajectory of the drop capsule
$\brhob'$, however there is a gravitational sag.

\subsubsection{Hamiltonian formulation of the canonical transformation from
  tower- to capsule coordinates}

In order to obtain the classical Hamiltonian of a single trapped particle
in capsule coordinates, we introduce the canonical transformation
\begin{equation}
  G(\brbp,\bp'',t)=\left[M\left(\brbp-\brhob'(t)\right)\right]\bp''
  +\frac{m}{\mathcal{M}}\left[\bar{s}'(t)+\brbp \boldsymbol{\bar{\pi}}'(t)\right],
\end{equation}
which yields
\begin{gather}
\br''=\frac{\partial G}{\partial \bp''}=M\left(\brbp-\brhob'(t)\right),\\
\bpbp=\frac{\partial G}{\partial \brbp}=M^{\top}(t)\bp''+\frac{m}{\mathcal{M}}\boldsymbol{\bar{\pi}}'(t).
\end{gather}

In this particular gauge, the one-particle Hamiltonian is given by
\begin{equation}
  \begin{aligned}
    \label{ham1}
    H_{sp}''&(\br'',\bp'',t) =\bar{H}_{sp}'(\brbp,\bpbp,t)+
    \bna{t}{} G(\brbp,\bp'',t)\\
    =&\frac{\bp''^2}{2\,m}+ m\widetilde{V}_t(\br'',\brho,t).
  \end{aligned}
\end{equation}
provided the equations for the capsule, Eqs. (\ref{hamroteq_1}) and (\ref{hamroteq_2}), 
are satisfied and $\bar{s}'(t,t_0)$ is an action,
\begin{gather}
\bar{s}'(t,t_0)=\int_{t_0}^t dt' \bar{L}_c'[\brhob',\dot{\brhob}',t']-\brhob'\,\boldsymbol{\bar{\pi}}'|^t_{t_0}.
\end{gather}
Obviously, this Hamiltonian is identical to Eq.~(\ref{hdprime}).


\subsection{Many identical atoms in the drop-capsule}

\subsubsection{Frame transformations between tower- and capsule 
coordinates}
The calculations of Sec.~\ref{secaccel} excluded the effects of rotation. In
here, we will rectify this and obtain a general frame transformation that
eliminates external forces and torques from Schr\"odinger's equation. In
particular, this is applied to the problem of gravitational acceleration and
Earth's rotation.

We have already introduced the generic many-particle Hamiltonian with
translationally invariant, binary interactions in \eq{inertialham}.  However,
we have not exploited the fact that only the s-partial wave contributes to the
two-particle scattering at low energies. Thus, we want to assume that the
relative angular momentum is a good quantum number. This can be modeled by an
interatomic potential that is only a function of the relative distance,
$V_p(|\bxi|)$. In the case of fermionic particles, this means there is no s-wave
interaction due to the Pauli exclusion principle and the gas becomes ideal at
low energies. 

Let us start with the full many-particle Hamiltonian 
\begin{gather}
  \label{ham2rot}
  \hat{\bar{H}}'=\hat{\bar{H}}'_{sp}(t)+\hat{\bar{V}}'_p,\\
  \hat{\bar{V}}'_p=\frac{1}{2}\int\,\text{d}^6 \bar{r}' \aopd{\brbp_1}
  \aopd{\brbp_2}  \bar{V}'_p(|\brbp_1-\brbp_2|)  \aop{\brbp_2} \aop{\brbp_1} \nonumber
\end{gather}
in the drop tower frame.

Now, we have to transform the Schr\"odinger state from the drop tower frame, $\ket{\Psi(t)}$,
to the rest frame of the drop capsule, $\ket{\Psi(t)}=\Uop(t)
\ket{\Psi(t)}'$. This requires to augment the unitary frame transformation from
\eq{norot} of Sec.~\ref{secdyn},
\begin{gather}
\label{Urot}
  \Uop(t)=\hat{U}_{\bar{\mathcal{S}'}(t)}\,\hat{U}_{\bar{\mathcal{P}'}(t)}\,
  \hat{U}_{\bar{\mathcal{R}'}(t)}\,\hat{U}^{\dagger}_{M(t)}, \\
  \intertext{with}
  \hat{U}_{M(t)}=e^{-i/\hbar\,\Qcb(t)\Lop}\,e^{-i/\hbar\,\phi\,\mathbf{e}_3\Lop}
                  \,e^{-i/\hbar\,\theta\mathbf{e}_2\Lop}.
\end{gather}
In there, we performed the same rotations as in Eq.~(\ref{M}), and in particular,
the diurnal rotation around the Earth axis, $\Qcb(t)=\Oeas\,t$, is accounted for.
Moreover, we would like to point out again that we transform from the drop tower frame, which is
corotating with the Earth, to the instantaneous rest frame of the drop capsule,
which is not rotating. Therefore we need to apply $\hat{U}^{\dagger}_{M(t)}$ rather 
than $\hat{U}_{M(t)}$.

The transformation rule for the Hamilton operator was given
in \eq{hop3} and consists of three contributions.  The first one contains merely
the gauge contributions that arise from the time-dependent frame parameters
\begin{equation}
  \begin{aligned}
    \Uopdag(t) (-i\hbar \frac{d}{dt})\Uop(t)=&(\dot{\bar{\Sc}}'+ \dot{\bar{\Pc}}'\,\bar{\Rc}')\,\Nop
    +
    \dot{\bar{\Pc}}' (M^{\top} \Rop)\\
    & - \dot{\bar{\Rc}}' (M^{\top} \Pop) +\Oeas \,\Lop.
\end{aligned}
\end{equation}
For obtaining this result, we have used the basic Eqs.~(\ref{nparticle}),
(\ref{displace}) and (\ref{rotate}).
The second contribution is the transformed single-particle Hamiltonian
\begin{equation}
  \begin{gathered}
    \Uopdag(t)\,\hat{\bar{H}}'_{sp}(t)\,\Uop(t)=\\
    \int \text{d}^3 r''\,\aopd{\br''} 
    \bar{H}'_{sp}(M^{\top}\br''+\bar{\Rc}', M^{\top}\bp''+\bar{\Pc}',t)
    \aop{\br''}.
  \end{gathered}  
\end{equation}
Finally, the third contribution is simply the transformed binary interaction
potential.
\begin{equation}
  \Uopdag(t)\, \hat{\bar{V}}'_p\,\Uop(t)=\hat{\bar{V}}'_p \equiv \hat{V}_p.
\end{equation}
It is left unchanged due to particle conservation, the local character of the
potential, as well as the translational and rotational invariance, \ie
\begin{equation}
 [\Nop,\hat{V}_p]=[\Rop,\hat{V}_p]= [\Pop,\hat{V}_p]=[\Lop,\hat{V}_p]=0.
\end{equation}

Combining these results, we find the transformed Hamiltonian in the
accelerated and non-rotating frame as
\begin{equation}
\label{accerotham}
\hat{H}''(t)= \int\,\text{d}^3r''\,
\aopd{\br''}\, H''_{sp}(\br'',\bp'',t)\, \aop{\br''}+\hat{V}_{p}.
\end{equation}
All gauge contributions are contained in the definition of the modified
single-particle Hamiltonian
\begin{equation}
  \label{transHrot}
  \begin{split}
    H''_{sp}(&\br'',\bp'',t)=
    \bar{H}_{sp}'(M^{\top} \br''+\bar{\Rc}',M^{\top} \bp''+\bar{\Pc}',t)\\
    &+\dot{\bar{\Sc}}' + \dot{\bar{\Pc}}'\,\bar{\Rc}' +\dot{\bar{\Pc}}'\,(M^{\top} \br'')
   - \dot{\bar{\Rc}}'\,(M^{\top} \bp'')\\
    &+\bp''\left(\Oeas\times\br''\right).
  \end{split}
\end{equation}


\subsubsection{Application of the frame transformation to the 
many-particle Hamiltonian in harmonic approximation}

We will now apply the considerations of the previous subsection
to the falling trapped
interacting many-particle system in the rotating frame. The single-particle
Hamilton operator follows from the classical considerations of
Eq.~(\ref{Hbarsp}) and is given by
\begin{align}
\bar{H}_{sp}'(&\brbp,\bpbp,t)=
\frac{\bpbp^2}{2 m}+V_t(M(t)(\brbp-\brhob'),t)\nonumber \\
&+V_g(M(t)(\brbp+\Reasb),t)-\bpbp\left(\Oeasb\times\brb\right).
\end{align}
By splitting the transformed Hamiltonian from Eq.~(\ref{transHrot}) into constant,
linear and higher order polynomial contributions in terms of the position
$\brbp$ and the momentum $\bpbp$, one finds
\begin{equation}
  \begin{split} 
    H''_{sp}(\br'',\bp'',t)=&
    \left[\boldsymbol{\alpha}\left(M^{\top}\bp''\right)+\boldsymbol{\beta}\left(M^{\top}\br''\right)
    +\gamma\right]\\
    &+\frac{\bp''^2}{2m}+m\widetilde{V}_{{t}}\left(\br'',M\bar{\Rc}'+\Reas,t\right).
\end{split}
\end{equation}
In there, we neglected terms of third and higher order in the gravitational potential.
The coefficients read 
\begin{align}
\boldsymbol{\alpha}=&\frac{\bar{\Pc}'}{m}-\dot{\bar{\Rc}}'-\Oeasb\times\left(\bar{\Rc}'+\Reasb\right),\\
\begin{split}
\boldsymbol{\beta}=&\dot{\bar{\Pc}}'-\bar{\Pc}'\times\Oeasb\\
&+m\,\bna{\bar{\Rc}'}{}[V_{t}(M(\bar{\Rc}'-\brhob'),t)+V_g(M(\bar{\Rc}'+\Reasb),t)].
\end{split}
\end{align}
and
\begin{equation}
    \gamma=\dot{\bar{\Sc}}'-\left[\bar{L}'_{sp}(\bar{\Rc}',\dot{\bar{\Rc}}',t)-\frac{d}{dt}\left(\bar{\Pc}'\bar{\Rc}'\right) \right].
\end{equation}
If the constraints are satisfied identically, \ie
$\{\boldsymbol{\alpha},\boldsymbol{\beta},\gamma\} \rightarrow 0$, one obtains
the single-particle Hamiltonian in the accelerated frame,
\begin{gather}
  \label{hpp}
  \begin{aligned}
    H_{sp}''(\br'',\bp'',t)= &\frac{\bp''^2}{2\,m}
    +m\,\widetilde{V}_{{t}}\left(\br'',M\bar{\Rc}'+\Reas,t\right),
  \end{aligned} 
 \end{gather}
 provided the frame parameters as well as the center-of-mass of the drop
 capsule  $\{\bar{\Rc}'(t),\bar{\Pc}'(t),\brhob'(t)\}$ satisfy the classical
 equations of motion 
\begin{gather}
  \label{piandpidot_rot}
  \bpib'=\bna{\dot{\brhob}'}{\bar{L}_c'}=
  \mathcal{M}\left(\dot{\brhob}'+\Oeasb\times\brhob
  \right),\quad  \dot{\bpib}'=\bna{\brhob'}{\bar{L}_c'},\\
  \label{cmosc_rot}
  \bar{\Pc}'=\bna{\dot{\bar{\Rc}}'}{\bar{L}_{sp}'}=
  m\left(\dot{\bar{\Rc}}'+\Oeasb\times(\bar{\Rc}'+\Reasb) \right),\quad
  \dot{\bar{\Pc}}'=\bna{\bar{\Rc}'}{\bar{L}_{sp}'}.
\end{gather}
With Eqs.~(\ref{piandpidot_rot}) and (\ref{cmosc_rot}), we have recovered the
Lagrangian $L'[\brhob',\dot{\brhob}',\bar{\Rc}',\dot{\bar{\Rc}}',t]$ of
Eq.~(\ref{lbarp}), and one can express the accrued dynamical phase as an
action integral
\begin{gather}
\bar{\Sc}'(t,t_0)=\int_{t_0}^t dt' \bar{L}_{sp}'[\bar{\Rc}',\dot{\bar{\Rc}}',t']-\bar{\Pc}'\bar{\Rc}'|^t_{t_0}.
\end{gather}


\section{Two-component atomic gas}
Let us now consider identical atoms with two internal electronic degrees of freedom. 
In this section, we would like to demonstrate that the dynamics of these internal states
decouples from the external motion of the trapped falling ultra-cold
quantum gas, even in case of binary collisions, if the self-scattering and the
cross-component scattering properties are identical.

\subsection{Description in second quantization}

Henceforth, the inclusion of two degrees of freedom for the inner states in
the definition of the field operators is required. These operators are denoted
by $\hat{a}_{\mu}(\br)$, $\mu \in \{1,2\}$, and satisfy the commutation
relation
\begin{equation}
\label{aadagger_2}
[\hat{a}_{\mu}(\br_1),\hat{a}^{\dagger}_{\nu}(\mathbf{\br}_2)]_{\mp}=
\delta_{\mu\nu}\,
\delta(\br_1-\br_2).
\end{equation}
The pair potential $\hat{V}_p$, which is assumed to be local and exhibits
translational and rotational invariance, can be written as \cite{merzbacher70}
\begin{equation}
\begin{split}
  \hat{V}_p =& \frac{1}{2}\int \text{d}^6 r\sum_{\mu_1..\mu_4=1}^{2}\,V_p^{\mu_1\mu_2\mu_3\mu_4}(|\br_1-\br_2|)\\
  &\phantom{\frac{1}{2}\int}\hat{a}_{\mu_1}^{\dagger}(\br_1)\,
  \hat{a}_{\mu_2}^{\dagger}(\br_2) \, \hat{a}_{\mu_3}(\br_2) \,
  \hat{a}_{\mu_4}(\br_1).
\end{split}
\end{equation}
The elements of $V_p^{\mu_1\mu_2\mu_3\mu_4}$ are linked to the self-species
and cross-component scattering lengths of the atom in the different states.
In $^{87}$Rb these quantities are -- to a good approximation -- all equal,
with a deviation of 3-4$\%$ \cite{matthews99-2}, so we can neglect the higher 
multipole contributions beyond the monopole (J=0) term
$V_p^{\mu_1\mu_2\mu_3\mu_4} \approx
\delta_{\mu_1 \mu_3}
\delta_{\mu_2 \mu_4}
\delta_{\mu_1 \mu_2} V_p^{(J=0)}$. This
leaves us with writing $\hat{V}_p$ approximately as
\begin{equation}
\begin{split}
  \hat{V}_p&=\frac{1}{2}\int \text{d}^6
  r\sum_{\mu,\nu}\hat{a}_{\mu}^{\dagger}(\br_1)\,
  \hat{a}_{\nu}^{\dagger}(\br_2) \,
  V_p^{(0)}(|\br_1-\br_2|)  \, \hat{a}_{\nu}(\br_2) \, \hat{a}_{\mu}(\br_1)\\
  &=\frac{1}{2}\int\text{d}^6r\,\hat{n}(\br_1) 
 (\hat{n}(\br_2)-\delta(\br_1-\br_2)) 
\, V_p^{(0)}(|\br_1-\br_2|).
\end{split}
\end{equation}
We made use of Eq. (\ref{aadagger_2}), and introduced the total particle
density
\begin{equation}
\hat{n}(\br)\equiv\sum_{\mu=1}^{2}\hat{a}_{\mu}^{\dagger}(\br) \, \hat{a}_{\mu}^{\phantom{\dagger}}(\br).
\end{equation}
We would like to point out that due to our assumptions $\hat{V}_p$ is now 
SU(2) invariant \cite{kuklov}.

The two internal states can be coupled by a classical, traveling-wave laser
field via electric dipole interaction \cite{schleich01}. The strength of the
coupling is determined by the Rabi frequency $\Omega(t)$, which may be
time-dependent if the laser is pulsed. $\Delta(t)$ denotes the, possibly
time-dependent, detuning of the laser with respect to the resonant transition
between the two levels. We want to consider large detunings in order to
neglect any mechanical recoil effects of the laser on the atoms and, as usual,
we have switched to an interaction picture oscillating with the laser
frequency \cite{schleich01}.

In order to describe the dipole interaction
of identical particles in the language of second quantization, we introduce
the operators
\begin{equation}
\hat{S}_i = \sum_{\mu,\nu=1}^{2}\,\int \text{d}^3 r \, \hat{a}_{\mu}^{\dagger}(\br)\,s^{(i)}_{\mu\nu}\,
 \hat{a}_{\nu}(\br)
\end{equation}
with the well-known spin-1/2-matrices $s^{(i)}$, $i \in \{ 1,2,3 \}$,
Appendix \ref{secb}.  Explicitly, the operators read
\begin{align}
  \hat{S}_1 &= \frac{\hbar}{2} \int \text{d}^3 r 
  \left[ \hat{a}_{2}^{\dagger}(\br)\hat{a}_{1}^{\phantom{\dagger}}(\br) 
    + \hat{a}_{1}^{\dagger}(\br)\hat{a}_{2}^{\phantom{\dagger}}(\br) \right],\\ 
\hat{S}_2&= i\frac{\hbar}{2} \int  \text{d}^3 r
\left[ \hat{a}_{2}^{\dagger}(\br)\hat{a}_{1}^{\phantom{\dagger}}(\br)
  -\hat{a}_{1}^{\dagger}(\br)\hat{a}_{2}^{\phantom{\dagger}}(\br) \right],\\
\hat{S}_3&= \frac{\hbar}{2}\int  \text{d}^3 r  
\left[ \hat{a}_{1}^{\dagger}(\br)\hat{a}_{1}^{\phantom{\dagger}}(\br)
  - \hat{a}_{2}^{\dagger}(\br)\hat{a}_{2}^{\phantom{\dagger}}(\br) \right].
\end{align}
These operators fulfill the commutation relations of the angular momentum
operators for spin-1/2-particles and therefore represent the SU(2) symmetry.
In other words, the properties of the Pauli matrices translate to the picture
of the second quantization. Clearly, the angular momentum algebra
\begin{equation}
\left[ \hat{S}_i, \hat{S}_j \right] = i \varepsilon_{ijk} \hat{S}_k
\end{equation}
holds.  All the operators $\mathbf{\hat{R}}$, $\mathbf{\hat{P}}$ and
$\mathbf{\hat{L}}$ commute with $\hat{S_{i}}$, i.e.
\begin{equation}
  \label{JR}
  \left[ \hat{S}_{i}, \hat{N} \right]=\left[ \hat{S}_{i}, \mathbf{\hat{R}} \right]=\left[ \hat{S}_{i}, \mathbf{\hat{P}} \right]
  =\left[ \hat{S}_{i}, \mathbf{\hat{L}} \right]=0.
\end{equation}
Please note, that the definition of operators $\mathbf{\hat{N}}$,
$\mathbf{\hat{R}}$, $\mathbf{\hat{P}}$, and $\mathbf{\hat{L}}$ has to be
modified, taking into account the two different internal states, i.e.
\begin{equation}
\hat{A}=\sum_{\mu=1,2}\int\text{d}^3r\,\hat{a}_{\mu}^{\dagger}(\br)\,A\,\hat{a}_{\mu}(\br),\end{equation}
where $\hat{A}\in\{\hat{N},\hat{\mathbf{R}},\hat{\mathbf{P}},\hat{\mathbf{L}} \}$, and 
$A\in\{1,\br,\mathbf{p},\br\times\mathbf{p}\}$.

We are now able to write down the full many-body Hamiltonian, containing the
one-particle, external dynamics ($\hat{H}_{sp}$), the one-particle internal
two-level dynamics ($\hat{V}_{d}$), describing the interaction between matter
and light, and the two-particle collisions ($\hat{V}_p$),
\begin{equation}
\label{totH}
\hat{H}=(\hat{H}_{sp}+\hat{V}_d)+\hat{V}_p.
\end{equation}
In there, 
\begin{equation}
\hat{H}_{sp}=\sum_{\mu=1,2}\,\int\text{d}^3 r\,\hat{a}^{\dagger}_{\mu}(\br)\,
H_{sp}\,\hat{a}_{\mu}(\br).
\end{equation}
$H_{sp}$ can be chosen as in Eq.~(\ref{Hbarsp}). $\hat{V}_d$ is given by
\begin{equation}
\label{h3}
\hat{V}_{d}=\hat{V}_d(t)=\Omega(t)\hat{S}_1 + \Delta(t) \hat{S}_3.
\end{equation}

\subsection{Separation of two-level dynamics and center-of-mass motion}

The dynamics of the many-particle system is determined by the Schr\"odinger equation 
\begin{equation}
\label{heisenberg}
i \hbar \partial_t \ket{\Psi(t)}
= \hat{H}\,\ket{\Psi(t)}
\end{equation}
with $\hat{H}$ from Eq.~(\ref{totH}).
As we have seen in the preceeding sections, a more efficient description of the dynamics
can be obtained by performing a frame transformation. In the present case, we have to
account for both external and internal dynamics. Therefore, it would be
favorable to separate both types of motion from each other.
We choose
\begin{equation}
\ket{\Psi(t)}=\hat{U}_{d}\,\hat{U}_{\bar{\mathcal{S}}'(t)}\,\hat{U}_{\bar{\mathcal{P}}'(t)}\,
  \hat{U}_{\bar{\mathcal{R}}'(t)}\,\hat{U}_{M(t)}^{\dagger}\ket{\Psi(t)}',
\end{equation}
cf. Eq.~(\ref{Urot}), and
\begin{equation}
\label{uj}
\hat{U}_{d} = \mathcal{T} \, \left\{ \, \exp\left[-\frac{i}{\hbar} \, \int_{t_0}^{t} \, dt'\, \hat{V}_d(t') \right] \right\},
\end{equation}
where $\mathcal{T}$ guarantees time ordering. The latter transformation formally
eliminates the two-level dynamics, since it cancels the contribution $\hat{V}_d$.
The decoupling of the transformation $\hat{U}_d$ from the remaining frame transformation
results from the fact that 
\begin{equation}
\left[\hat{V}_{d},\hat{H}_{sp}+\hat{V}_p\right]=0
\end{equation}
due to Eq.~(\ref{JR}) as well as the SU(2)-invariance of $\hat{V_p}$, cf.
preceeding subsection.  In principle, the propagator $\hat{U}_{d}$ can be
determined from the two linearly independent solutions of the time-dependent
Rabi problem \cite{eberly, barata00, bagrov01}.

\section{Conclusions}
We have described the dynamics of an ultra-cold quantum gas in a long distance
free-fall experiment.  Starting from the classical mechanics of the drop
capsule and a single trapped particle, we developed the quantum-field
theoretical description of a trapped, interacting degenerate quantum gas in a
drop experiment in an inertial frame, the corotating frame of the Earth and
the comoving frame of the drop capsule. By introducing suitable coordinate
transformations, it was possible to eliminate non-inertial forces and to focus
on effects that take place on the mesoscopic length scale of the Bose or Fermi
gas.  The exact cancellation of non-inertial forces requires translational
invariance, the isotropy of the binary collisional potential and the presence
of a quadratic single-particle Hamiltonian.  This is well satisfied for
$^{87}$Rb, and the harmonic approximation of the gravitational potential
around the center-of mass of the BEC wave-packet is an excellent assumption.
Corrections to it could be easily calculated perturbatively.

If the atoms are two-level systems and coupled by an off-resonant
traveling-wave laser field, this internal dynamics can be separated from the
external motion, provided all scattering lengths are identical
(SU(2)-invariance).

This formalism provides us with an efficient way to describe free-fall
experiments, especially for numerical studies. It almost goes without saying
that it is in particular valid and useful on the mean-field level.

While we have discussed the Euclidean transformations corresponding to
translation and rotation, we have omitted the scaling transformations
\cite{pitaevskii97}, which are obviously very useful to model the adiabatic
expansion of a BEC from a trap \cite{castin}.  The consideration of
the complete set of generalized canonical transformations in a time-dependent
way will ultimately separate all "trivial" dynamics, including wave-packet
spreading, from the essential many-particle physics. This is work in progress.
We have also not touched questions of relativity. This was done by intention
in order to clarify all non-relativistic effects first, which by themselves
are highly nontrivial and presumably dominant.


\section*{Acknowledgments}
We thank the members of the QUANTUS collaboration \cite{quantus} A. Vogel, K.
Sengstock, K. Bongs (Institut f\"ur Laser-Physik, Universit\"at Hamburg),
W.~Lewoczko, M.~Schmidt, T.~Schuldt, A.~Peters (Institut f\"ur Physik,
Humboldt-Universit\"at zu Berlin), T.~van Zoest, T.~K\"onemann, W.~Ertmer,
E.~Rasel (Institut f\"ur Quantenoptik, Universit\"at Hannover), T.~Steinmetz,
J.~Reichel (Laboratoire Kastler Brossel de l'E.N.S.), W. Brinkmann,
E.~G\"okl\"u, C.  L\"ammerzahl, H.J.~Dittus (ZARM University of Bremen) for
the fruitful collaboration and the DLR (Deutsches Zentrum f\"ur Luft- und
Raumfahrt) for the financial support of the project {\em Quantum Gases in
  Weightlessness} (DLR 50 WM 0346).

\appendix
\section{Taylor expansion of the gravitational potential}
\label{apptay}
The Taylor expansion of the gravitational potential
\begin{equation}
\begin{gathered}
  V_g(\mathbf{r'}+\brho,t)=[1+\mathbf{r'}\,\bna{\brho}{}+
  \frac{1}{2!}(\mathbf{r'}\,\bna{\brho}{})^2+\ldots]\,V_g(\brho,t)\\
  =V_g(\brho,t)+V_g^{(1)}(\br',\brho,t)+V_g^{(2)}(\br',\brho,t)\\
  +\delta V_g^{(3)}(\br',\brho,t).
\end{gathered}  
\end{equation}
around the center-of-mass of the wavepacket defines a gradient field,
Eq.~(\ref{grad}), a symmetric Hessian tensor, Eq.~(\ref{hess})
and the remainder forms a residual potential
\begin{gather}
  \delta V_g^{(3)}(\br',\brho)=\sum_{i=3}^{\infty} V_g^{(i)}(\br',\brho),\\
  \intertext{ which starts off with a leading third-order correction as}
  \begin{aligned}
    &V_g^{(3)}( \br',\brho)=\frac{1}{3!} \,(\br'\otimes \br' \otimes\br')
    \cdot v_g^{(3)}(\brho),\\
    &v_g^{(3)}(\brho)=\bna{\brho}{} \otimes \bna{\brho}{} \otimes\bna{\brho}{}
    \ V_g(\brho).
  \end{aligned}
\end{gather}

In order to obtain estimates for the Taylor coefficients at
the surface of the Earth, we use the isotropic gravitational potential and its
derivatives
\begin{gather}
  V_g(\br)=-\frac{G M_\eas}{r},\quad
  \frac{d^n}{dr^n} V_g=(-1)^{n+1}n!\,\frac{G M_\eas}{r^{n+1}}.
\end{gather}
Now, if this potential is expanded around a point $\brho$, we obtain the
well-known multipole expansion in terms of scalar, dipolar and quadrupolar
components
\begin{equation}
\begin{gathered}
V_g(r)=V_g\left(|\brp+\brho|\right)=
V_g(\rho)+\brp \frac{\brho}{\rho} \frac{dV_g}{d\rho}\\+
\frac{1}{2}(\brp\otimes\brp)
\left[\mathds{1}\frac{1}{\rho}\frac{dV_g}{d\rho}
+\frac{\brho\otimes \brho}{\rho^2}(\frac{d^2V_g}{d\rho^2}
-\frac{1}{\rho}\frac{dV_g}{d\rho})\right]\\
=V_g(\rho)\left(1-\frac{\brp\brho}{\rho^2}+
\frac{1}{2}(\brp\otimes\brp)\left[
3\frac{\brho\otimes \brho}{\rho^4}-\frac{\mathds{1}}{\rho^2}\right]\right).
\end{gathered}
\end{equation}
\section{Useful relations for the construction of the 
  Lie groups elements}
\label{secb}
\subsection{Euclidean space}
The angular momentum matrices ($L=1$) $\ell_{i=1,2,3}$ satisfy the angular
momentum algebra of Eq.~(\ref{angmomentum}). In a Cartesian basis, the matrix
representation is $[\ell_l]_{jk}=-i \hbar\,\epsilon_{ljk}$ or explicitly
\begin{gather}
  \ell_1=\hbar
  \begin{pmatrix}
    0&0&0\\
    0&0&-i\\
    0&i&0
  \end{pmatrix}, \quad
  \ell_2=\hbar
  \begin{pmatrix}
    0&0&i\\
    0&0&0\\
    -i&0&0
  \end{pmatrix},\\
  \ell_3=\hbar
  \begin{pmatrix}
    0&-i&0\\
    i&0&0\\
    0&0&0
  \end{pmatrix}.
\end{gather}
A finite orthogonal rotation is formed from these generators by
\begin{eqnarray}
  M_\Qc\,\br&=&e^{-i/\hbar\, \Qc \,\boldsymbol{\ell}}\ \br=\br+\Qc\times\br+ \ldots.
\end{eqnarray}

For $L=1/2$, the spin matrices $s^{(i)}$ satisfy the angular
momentum algebra of Eq.~(\ref{angmomentum}) as well.
They are given by
\begin{gather}
s^{(1)}=\frac{\hbar}{2}\sigma^{(1)}=\frac{\hbar}{2}
\begin{pmatrix}
0 & 1\\1 & 0
\end{pmatrix},\\
s^{(2)}=\frac{\hbar}{2}\sigma^{(2)}=\frac{\hbar}{2}
\begin{pmatrix}
0 & -i\\i & 0
\end{pmatrix},\\
s^{(3)}=\frac{\hbar}{2}\sigma^{(3)}=\frac{\hbar}{2}
\begin{pmatrix}
1 & 0\\0 & -1
\end{pmatrix}.
\end{gather}
In there, $\sigma^{(i)}$ are the Pauli spin matrices \cite{schleich01}.

\subsection{Single-particle Hilbert space}
If the unitary operators act on the Heisenberg position and momentum operator,
we find
\begin{xalignat}{2}
    U_\mathcal{R}^\dag\, \br\, U_\mathcal{R}^{\vphantom \dag}&=\br+\Rc,&\quad
    U_\mathcal{P}^\dag\, \bp\, U_\mathcal{P}^{\vphantom \dag}&=\bp+\Pc,\\
    U_\Qc^\dag\, \br\, U_\Qc^{\vphantom \dag}&=M_\Qc\, \br,&\quad
    U_\Qc^\dag\, \bp\, U_\Qc^{\vphantom \dag}&=M_\Qc\, \bp.
  \end{xalignat}
Please note that we have suppressed here extraneous hats to distinguish them from
ordinary vectors in Euclidean space.

\subsection{Many-particle Fock space}
In the course of our calculations, we make use of some more 
basic commutator relations
\begin{alignat}{2}
    \left[\aop{\br}, \Nop \right] &
    = \aop{\br},&\quad
    \left[\aop{\br}, \Rop \right] &
    = \br \, \aop{\br},\\
    \left[\aop{\br}, \Pop \right] &
    = \bp \, \aop{\br},&\quad
    \left[\aop{\br}, \Lop \right] &
    = \mathbf{l} \, \aop{\br}.
\end{alignat}
They induce the unitary representations of the translational and rotational group
\begin{xalignat}{2}
  \UN&=e^{i/\hbar\,\Sc\,\Nop}, &\quad 
  \UP&=e^{i/\hbar\,\Pc\,\Rop},\\
  \UR&=e^{-i/\hbar\,\Rc\,\Pop},&\quad 
  \UO&=e^{-i/\hbar\, \Qcb\, \Lop},
\end{xalignat}
and we obtain again a representation of the group operations
\begin{alignat}{2}
  \label{displace}
  \URdag \, \Rop\, \UR &= 
  \Rop + \Rc\,\hat{N}, &\quad
  \UPdag \, \Pop\, \UP &= 
  \Pop + \Pc\,\hat{N},\\
  \label{rotate}
  \UOdag\,\Rop \,\UO&=
  M_\Qc\,\Rop,&\quad
  \UOdag\,\Pop \,\UO&=M_\Qc\,\Pop.
\end{alignat}

\section{Classical trajectory of the drop capsule in a rotating frame}

In Section IV, the free-fall experiment in a rotating frame is discussed. A sketch of the physical situation is given in Fig.~\ref{coord}. 
$\theta$ and $\Theta$ are linked via
\begin{equation}
\tan\theta=\frac{b}{a}\tan\Theta.
\end{equation}

The classical trajectory of the drop capsule in the rotating frame of the
Earth is given by
\begin{gather}
  \rhob_x'(t) = \bar{g}\sin\theta\cos\theta\left(\frac{t^2}{2}
    -\frac{1-\cos(2\Oet)}{4\Oe^2}\right)\\
  \rhob_y'(t) = \bar{g}\frac{\sin\theta}{2\Oe}(t
  -\frac{\sin(2\Oet)}{2\Oe})\\
  \rhob_z'(t) = h-\frac{\bar{g}}{2}t^2+\bar{g}\sin^2\theta\left(\frac{t^2}{2}
    -\frac{1-\cos(2\Oet)}{4\Oe^2}\right) ,
\end{gather}
where $\Oe=|\Oeasb|$ and
$\bar{g}=|\mathbf{\bar{g}}|$.  $\bar{\brho}'(t)$ solves
Eq.~(\ref{classicnewlin}), if the capsule is released with the initial
conditions $\bar{\brho}'(t=0)=(0,0,h)^{\top}$,
$\dot{\brhob}'(t=0)=\mathbf{0}$, and if the centrifugal term
is neglected.

The rotation matrix $M_{\mathcal{Q}}$ evolves around the axis
$\Oeas$ with an angle $\mathcal{Q}=\Oet$ (see
Fig.~\ref{coord}). Explicitly, it is given by
\begin{equation}
M_{\mathcal{Q}} = \begin{pmatrix}
 \cos\mathcal{Q}\cos\theta& -\sin\mathcal{Q} & \cos\mathcal{Q}\sin\theta \\
 \sin\mathcal{Q}\cos\theta & \cos\mathcal{Q} & \sin\mathcal{Q}\sin\theta\\
 -\sin\theta & 0 & \cos\theta
\end{pmatrix}.
\end{equation}

\bibliography{ref_droptower,bec,MyPublications}

\end{document}